\documentstyle[prl,multicol,aps]{revtex}

\newcommand{\be}{\begin{equation}}
\newcommand{\ee}{\end{equation}}
\newcommand{\bey}{\begin{eqnarray}}
\newcommand{\eey}{\end{eqnarray}}

\begin{document} 
\draft
%\twocolumn[\hsize\textwidth\columnwidth\hsize\csname@twocolumnfalse\endcsname
%\preprint{HKBU-CNS-9815}

\title{ Perturbative and non-perturbative parts of eigenstates
and local spectral density of states: the Wigner band random matrix model
}

\author{Wen-ge Wang }
\address{ 
Department of Physics, South-east University, Nanjing 
210096, China 
\\ International Center for the Study of
Dynamical Systems, University of Milan at Como, 
via Lucini 3, 22100 Como, Italy 
\\ Department of Physics and Centre for Nonlinear Studies, 
Hong Kong Baptist University, Hong Kong, China 
}
 
\date{\today}
\maketitle

\begin{abstract} 
A generalization of Brillouin-Wigner perturbation theory is applied
numerically to the Wigner Band Random Matrix model. The perturbation
theory tells that a perturbed energy eigenstate can be divided into a perturbative part
and a non-perturbative part with the perturbative part expressed
as a perturbation expansion. Numerically it is found that such a
division is important in understanding many properties of
both eigenstates and the so-called local spectral density of states (LDOS).
For the average shape of eigenstates, its
central part is found to be composed of its non-perturbative part
and a region of its perturbative part, which is close to the non-perturbative
part. A relationship between the average shape of eigenstates
and that of LDOS can be explained. Numerical results also show that
the transition for the average shape of 
  LDOS from the Breit-Wigner form to the semicircle form
is related to a qualitative change in some properties of the perturbation
expansion of the perturbative parts of eigenstates. 
The transition for the half-width of the LDOS from quadratic dependence to linear dependence
on the perturbation strength is  accompanied
by a transition of a similar form for the average size of the non-perturbative
parts of eigenstates. For both transitions, the same critical 
perturbation strength $\lambda _b$ has been found to play important 
roles. When perturbation strength is larger than $\lambda _b$, the 
average shape of LDOS obeys an approximate scaling law. 
\end{abstract}
\pacs{PACS number 05.45.+b}

\begin{multicols}{2}

%\narrowtext

\section{Introduction}

The average shape of energy eigenfunctions (EFs),
which characterizes the spreading of the eigenstates
of a perturbed system over the eigenstates of an unperturbed system,
 is very important in a
wide range of physical fields, from nuclear physics and atomic physics to
condensed matter physics. 
For example, in a recent approach for the description of isolated
Fermi systems with finite number of particles, a type of ``microcanonical''
partition function has been introduced and expressed in terms of the average
shape of eigenfunctions \cite{FI97}. However, up to now only some of the general
features of the shape have been known clearly.
For example, when perturbation is not strong, generally the shape has a Breit-Wigner
form (Lorentzian distribution) \cite{GS97,JS95,FM95,Frahm95}, 
but in a larger range on the interaction strength, 
without any simple analytical expression known for the shape, 
only some phenomenological expressions
have been suggested  
\cite{FGGK94,ZBHF96,FGI96,FIC96,FI97R,CFI}.

Another quantity, the so-called local spectral density of
states (LDOS), has also attracted lots of attention recently
\cite{FGGK94,ZBHF96,FCIC96,CCGI96,JSS97,MF-cond,BGIC97}. This quantity,
known in nuclear physics as the ``strength function'', gives
information about the ``decay'' of a specific unperturbed state into
other states due to interaction. In particular, the width of the
LDOS defines the effective ``lifetime'' of the
unperturbed  state. For sufficiently weak coupling, the
average shape of LDOS is usually close to the Breit-Wigner form with
the half-width having a quadratic dependence on the interaction
strength. On the other hand, when  interaction is strong, the
shape is model dependent and the
half-width is linear in the interaction strength.

Numerically it is already known that for Hamiltonian 
matrices with band structure generally both the average shape
of EFs and that of LDOS can be divided into two parts: central parts and
tails with exponential (or faster) decay. However, an analytical
definition for such a division has not been achieved yet.
A possible clue for this problem comes from a generalization
of the so-called Brillouin-Wigner perturbation theory \cite {Ziman}
introduced in Ref. \cite{WIC98} for studying long tails of EFs, which tells that analytically EFs can be divided
into perturbative and non-perturbative parts with perturbative
parts being able to be expanded in a perturbation expansion
by making use of the corresponding non-perturbative parts.
The relationship between central parts and non-perturbative
parts of EFs was not studied in Ref. \cite{WIC98}, while
it is more important.

The purpose of this paper is to study if the above mentioned
generalization of Brillouin-Wigner perturbation theory (GBWPT)
can provide deeper understanding for properties of EFs and LDOS,
especially the relationship between the two divisions
mentioned above for EFs. As a first step, here we study the so-called
Wigner band random matrix (WBRM) model, which is one of the simplest
models for the application of the GBWPT. 

The WBRM was first introduced by Wigner 40 years ago \cite{Wigner}
for the description of complex quantum systems as nuclei. It is
currently under close investigation
\cite{FGGK94,FCIC96,CCGI96,FLP89,FLW91,LF93,CCGI93,GFIM93,PR93}
since it is believed to provide an adequate description also for
some other complex systems (atoms, clusters, etc.) and as well
as for dynamical  conservative systems with few degrees of
freedom, which are chaotic in the classical limit. Unlike the
standard Band Random Matrices (see, for example, \cite{FM94,Izrailev95}),
the theory of WBRM is not well developed. Numerically it is known
that different averaging procedures for EFs give different results
for their average shape. Specifically, when averaging is done with
respect to centers of energy shell, for strong interaction  
central parts of averaged  EFs are close to
a form predicted for the LDOS by the semicircle
law. On the other hand, for the case of averaging
with respect to centroids of EFs, central parts of averaged EFs
are obviously narrower than the corresponding LDOS when the so-called
Wigner parameter is large and an ergodicity parameter is small
\cite{CCGI96}.

Comparatively, more properties are known for the average shape of the LDOS of the  
WBRM. For the tails, a corrected analytical expression has been
derived in Ref. \cite{FGGK94} for weak perturbation and numerically
found correct for even strong perturbation \cite{FGGK94,CFI,FCIC96}.
A more general analytical approach for the LDOS of the WBRM is given
in Ref. \cite{FCIC96}, where an expression for the LDOS is given 
in terms of a function satisfying an integral equation.
In sufficiently weak and extremely strong
perturbation cases, the expression gives  the well known Breit-Wigner
form and semicircle law, respectively.
For the transition from the Breit-Wigner form to the semicircle form,
although the expression also predicts correct results for the LDOS,
the physical explanation for it is not so clear yet. 
Therefore, it is  of interest to study if the GBWPT can 
throw new light on the above mentioned properties of the 
EFs and the LDOS of the WBRM. 

This paper has the following structure. 
For the sake of completeness, in section II
we  give a brief presentation of the GBWPT. Some predictions
for properties of EFs of the WBRM are also given in the section. Numerical
results for the size of non-perturbative parts of EFs are
discussed in section III. It is shown that for the average shape of
EFs the averaged non-perturbative part is indeed related to the central part. A region
predicted in section II, which is between the non-perturbative part
and the (exponentially or faster decaying) tails of an EF, is also studied
numerically. Section IV is devoted to a discussion for the
relation between the average shape of EFs and that of the 
LDOS. In section V, some properties of the average 
shape of LDOS, such as the transition from the Breit-Wigner 
form to the semicircle form and the dependence of its half-width 
on perturbation strength, are studied numerically and found to be closely 
related to properties of the average size of the non-perturbative 
parts of the corresponding EFs. Finally, conclusions are given in section VI.

\section{Generalization of Brillouin-Wigner perturbation theory}

For the sake of completeness, in this section
we first give a brief presentation of  the above mentioned 
generalization of Brillouin-Wigner perturbation
theory (GBWPT) \cite{WIC98}.  Then, we  give a brief discussion 
for some properties of the EFs of the WBRM.

Generally, consider a Hamiltonian of the form 
\be \label{H}  H=H^0 + \lambda V \ee
where $H^0$ is an unperturbed Hamiltonian, $V$ is a perturbation and
the parameter $\lambda $ is for adjusting the strength of the perturbation. 
For the WBRM with dimension $N$ and band width $b$, 
the   Hamiltonian matrix considered here is chosen of the form 
\be 
H_{ij} = (H^0 + \lambda V)_{ij} = 
E^0_i \delta _{ij} + \lambda v_{ij}  
\label{Hij} \ee
where
\be \label{E0} E^0_i =i \ \ \ \ \ \ (i=1, \cdots , N) \ee
are eigenenergies of the eigenstates of $H^0$ labeled by $|i\rangle $. 
Off-diagonal matrix elements $v_{ij} = v_{ji}$ 
are random numbers with Gaussian distribution for 
$1 \le |i-j| \le b$ ($\langle v_{ij} \rangle = 0$
and $\langle v^2_{ij} \rangle =1$)
and are zero otherwise.  
Here basis states have been chosen as  
the  eigenstates $|i \rangle $ of $H^0$ in energy 
order.  
Eigenstates of $H$, labeled by $|\alpha \rangle$, are also ordered 
in energy, 
\be \label{Ea} 
H |\alpha \rangle = E_{\alpha} |\alpha \rangle,
\ \ \ \ \ \ E_{\alpha +1} \ge E_{\alpha }. \ee 

In order to obtain the GBWPT, let us divide the 
set of basis states  
 into two parts, $\{|i \rangle, \ i=p_1, p_1+1, \cdots 
p_2 \}$ and $\{ |j \rangle, \ j=1, \cdots p_1-1,p_2+1, 
\cdots N \}$. This also divides the Hilbert space into
two sub-spaces, for which the corresponding 
projection operators are 
\be P \equiv \sum_{i=p_1}^{p_2} |i \rangle \langle i| 
\ \ \ \ and \ \ \ \ Q \equiv 1-P. 
\label{PQ} \ee
Subspaces related to the projection operators $P$ and $Q$ will be called 
in the following the {\it P} and 
{\it Q subspaces }, respectively.
 For an arbitrary eigenstate $|\alpha \rangle $, which is split  
 into two orthogonal parts
$|t \rangle \equiv P|\alpha \rangle$ and $|f \rangle
\equiv Q |\alpha \rangle$ by the projection operators $P$ and $Q$,  
by making use of the stationary Schr\"{o}dinger equation, it can be shown that 
\be |\alpha \rangle = |t \rangle + \frac 1{E_{\alpha} - H^0} 
Q\lambda V|\alpha \rangle.  
\label{GBW} \ee
Introducing an operator $T$,
\be T \equiv \frac 1{E_{\alpha}-H^0}Q\lambda V,
\label{T} \ee 
the iterative expansion of Eq.~(\ref{GBW}) is of the form  
\be |\alpha \rangle = |t \rangle +   T|t \rangle
+ T^2|t \rangle  + \cdots+ T^{n-1}|t \rangle
+ T^n |\alpha \rangle . \label{GBWE} \ee
Then, one can see that if the projection operators $P$ and $Q$
are such chosen that
\be \lim _{n \to \infty }  \langle \alpha |(T^{\dagger})^n
T^n |\alpha  \rangle =0,  \label{TnTn} \ee
Eq.~(\ref{GBWE}) gives  
\be |\alpha \rangle = |t \rangle +
T|t\rangle
+ T^2|t\rangle
+ \cdots + T^n|t\rangle + \cdots.
 \label{e-psi} \ee
 Here the eigenvalue $E_{\alpha }$ has 
been treated as a constant. Eq.~(\ref{e-psi}) is a
 {\it generalization } of the so-called 
{\it    Brillouin-Wigner perturbation 
expansion (GBWPE)} (for BWPE, see, for example, \cite{Ziman,XWwgY92}).

Equation (\ref{TnTn}) is the condition for the GBWPE (\ref{e-psi}) to hold. 
Generally, if $P$ and $Q$ subspaces are such chosen that
$|E_{\alpha} -E^0_j|$ is large enough compared with $b \lambda V$
for any basis state $|j \rangle$ in the $Q$ subspace,
 $|T_{\alpha}^n>$ will
vanish when $n \to \infty$. 
Therefore, generally there are many choices for the $P$ and $Q$ 
operators ensuring Eq.~(\ref{e-psi}) to hold. Among them, 
the projection operator(s)  $P$  with the minimum
number of basis states and the
corresponding operator(s) $Q$ with the maximum number of basis states  
will be denoted by $P_{\alpha }$
and $Q_{\alpha }$, respectively, and the corresponding 
 $|t \rangle $
and $|f \rangle $ parts of the state $|\alpha \rangle $ will
be denoted by 
$|t_{\alpha } \rangle \equiv P_{\alpha }|\alpha \rangle$ 
and  $|f_{\alpha }\rangle \equiv Q_{\alpha }|\alpha \rangle $, respectively. 
The $|t_{\alpha }\rangle $ part of the state $|\alpha \rangle $
will be called the {\it non-perturbative } (NPT) part of 
$|\alpha \rangle$, and the $|f _{\alpha }\rangle $ part called the
{\it perturbative } (PT) part of $|\alpha \rangle$.
Correspondingly, the eigenfunction of the state $|\alpha \rangle $, i.e.,
its components in the basis states, can also be divided into
NPT and PT parts. 
Eq.~(\ref{e-psi}) tells that $|f_{\alpha } \rangle$
can be expressed in terms of $|t_{\alpha}\rangle$, $E_{\alpha }$, 
$\lambda V$ and $H^0$.
In what follows, generally 
the GBWPE (\ref{e-psi}) will be used to  
discuss the perturbative parts of eigenstates only. 

To achieve a more explicit condition for Eq.~(\ref{TnTn}) to hold,
let us write $T^n $ in the following form,
\be T^n  = \frac 1{E_{\alpha }-H^0}
\cdot (\lambda U)^{n-1} \cdot Q\lambda V,
\label{TU} \ee
where
\be U \equiv Q V \frac{1}{E_{\alpha }-H^0} Q
\label{U} \ee
is an operator in the $Q$ subspace. 
Eq.~(\ref{TU}) shows that it is the properties of $\lambda U$ that
determines if $T^n |{\alpha } \rangle $ vanishes when $n$ approaches to 
$\infty $. Indeed, writing eigenstates of the operator $U$ in the $Q$
subspace as $|\nu \rangle $, 
\be U |\nu \rangle = u_{\nu } |\nu \rangle , \label{Uu} \ee
it is easy to see that if all the values of $|\lambda u_{\nu} |$
are less than 1, then
$T^n |{\alpha } \rangle $ vanishes when $n$ goes to infinity.
Therefore, generally the condition for Eq.~(\ref{e-psi}) (also
Eq.~(\ref{TnTn})) to hold is that the corresponding $\lambda U$ operator
in the $Q$ subspace does not have any eigenvector $|\nu \rangle $ with
 $|\lambda u_{\nu }| \ge 1$. 

As an application of the above results to the WBRM, let us study
the component of an eigenstate $|\alpha  \rangle $ on a basis
state $|j \rangle $ in the $Q_{\alpha }$ subspace,
$C_{\alpha j} \equiv \langle j|\alpha \rangle $, i.e., a component
of the perturbative part $|f_{\alpha } \rangle $. 
In what follows, for convenience, we use
$|j \rangle $ to indicate a basis state in the $Q_{\alpha }$
subspace, $|i \rangle $ to indicate a basis state in the $P_{\alpha }$
subspace and $|k \rangle $ for the general case. 
Suppose $j$ is larger than $p_2$ for the $P_{
\alpha }$ subspace, specifically, $p_2+mb < j \le p_2+(m+1)b$ with
$m$ being an integer greater than or equal to zero.  Noticing that
a result of the band structure of the Hamiltonian matrix of the
WBRM is that 
$\langle j|(QV)^n|t_{\alpha }\rangle$ is zero when $n$ is less than
$(m+1)$, from Eqs.~(\ref{e-psi}) and (\ref{TU}) we have
\bey \label{CajT}  C_{\alpha j} = \langle j| (T^{m+1} + T^{m+2}
+ \cdots ) |t_{\alpha } \rangle \ \ \ \ \ \ \ \ \ \ \ \ \ \ \ \ \ \ 
\nonumber \\ 
 = \langle j| \frac 1{E_{\alpha }-H^0}(\lambda U)^m
 (1+\lambda U + \cdots )
Q\lambda V |t_{\alpha} \rangle . \eey 
Since $Q\lambda V|t_{\alpha }\rangle $ is a vector in the
$Q_{\alpha }$ subspace, it can be expanded in the eigenstates
of $U$, 
\be Q\lambda V |t_{\alpha }\rangle = \sum_{\nu } h_{\nu }
|\nu \rangle . \label{QVnu} \ee
A result of this expansion and Eq.~(\ref{Uu}) is 
\bey \label{Uhh} (1+\lambda U + (\lambda U)^2 + \cdots )Q\lambda V
|t_{\alpha }\rangle \ \ \ \  \ \ \  \nonumber \\
= \sum_{\nu } h_{\nu } (1+\lambda u_{\nu } +
(\lambda u_{\nu })^2 + \cdots ) |\nu \rangle \nonumber \\
= \sum_{\nu } \frac{h_{\nu }}{1-\lambda u_{\nu }}
|\nu \rangle . \ \ \ \ \ \ \ \ \ \ \ \ \ \ \ \ \ \ \ \ \ \ \  \eey
Then, substituting Eq.~(\ref{Uhh}) into Eq.~(\ref{CajT}), one has 
\be C_{\alpha j} = \frac 1{E_{\alpha }-E^0_j} \sum_{\nu }
[ \frac {h_{\nu }}{1- \lambda u_{\nu}} \langle j|\nu \rangle ]
(\lambda u_{\nu })^m . \label{CU} \ee
Since $|\lambda u_{\nu}| <1$ for all $\nu $, the behavior of
the long tails of the EFs of the WBRM is more or less like exponential
decay as discussed in Ref. \cite{FGGK94}.
In fact, from the viewpoint of the GBWPT, the proof and arguments
given in Ref. \cite{FGGK94} for behaviors of long tails of the LDOS
of the WBRM are still valid when perturbation is strong.

What is of special interest here is the behavior of $C_{\alpha j}$ for
$j$ in the regions of $(p_2, p_2+b]$ and $[p_1-b,p_1)$, 
i.e., for $j$ in the regions of size $b$ just beside the non-perturbative
part of the eigenstate.  For these $j$,
  $m=0$ in Eq.~(\ref{CU}) and it is not necessary for
  $|C_{\alpha j}|$ to decay exponentially. That is to say, 
the decaying speed of 
$|C_{\alpha j}|$ for these $j$ should be slower than 
that for $j$ larger than $ p_2+b$ or less than $p_1-b$. 
The two regions of  $j \in (p_2,
p_2+b]$ and $j \in [p_1-b,p_1)$ will be called the {\it slope} regions of the eigenstate $|\alpha \rangle $ 
( the reason for such a name will be 
clear from numerical results in the next section). 

According to Eq.~(\ref{e-psi}), 
the explicit expression for $C_{\alpha j}$ in terms of elements of
the Hamiltonian matrix can be written as 
\bey C_{\alpha j}  = 
\sum_{i=p_1}^{p_2} ( \frac {\lambda V_{ji}}{E_{\alpha} - E^0_j} 
+ \sum_{k \in Q} \frac {\lambda V_{jk}}{E_{\alpha} - E^0_j} 
\frac {\lambda V_{ki}}{E_{\alpha} - E^0_k}  
\nonumber \\
+ \sum_{k,l \in Q} \frac {\lambda V_{jk}}{E_{\alpha} - E^0_j} 
\frac {\lambda V_{kl}}{E_{\alpha} - E^0_k} 
\frac {\lambda V_{li}}{E_{\alpha} - E^0_l} \cdots )
C_{\alpha i}   \label{cja} \eey 
where $C_{\alpha i} = \langle i|t_{\alpha } \rangle $.
This expression can be written in a simpler form  by
making use of the concept of path, in analogy to that in  the 
  Feynman's path integral theory \cite{F65}, which can be
  done as follows.
For $(q+1)$ basis states $\{ |j\rangle , |k_1\rangle , \cdots |k_{q-1}\rangle , 
|i\rangle \}$ with only $|i\rangle $ in the $P_{\alpha }$ subspace, 
 we  term the sequence  $j \to k_1 \to 
\cdots \to k_{q-1} \to i$ 
{\it a path of q paces} from $j$ to $i$,
if $V_{kk'}$ corresponding to  each pace is non-zero. 
Clearly, paths from $j$ to 
$i$ are determined by the structure of the Hamiltonian matrix in 
$H^0$ representation. 
Attributing a  factor $V_{kk'}/(E_{\alpha} -E^0_k)$ to each 
pace $k \to k'$, 
 the contribution of a path $s$  to $C_{\alpha j}$, 
denoted by $f_{\alpha s}(j \to i)$,   
is defined as the product of  the  factors of all its  paces.  
  Then,  $C_{\alpha j}$ in Eq.~(\ref{cja})  
 can be rewritten as 
\begin{equation}
C_{\alpha j} = \sum_{i=p_1}^{p_2} \sum_s f_{\alpha s}
(j \to i) C_{\alpha i}.   
\label{Cf} \end{equation}

An advantage of expressing $C_{\alpha j}$ in terms of  
contributions of paths is that it makes it easier 
to understand the important  role played by 
the size of the non-perturbative part of the EF 
in determining its shape. Such a size for the state $|\alpha
\rangle $ will be denoted by $N_p $, 
$N_p \equiv p_2-p_1+1$. ($N_p$ is, in fact, a function of $\alpha $,
but, for brevity, the subscript $\alpha $ will be omitted.)
Two values of $\lambda $   are of special interest
when $\lambda $ increases from zero.
One is the smallest $\lambda $ for $N_p =2$, indicating the beginning
of the invalidity of 
the ordinary Brillouin-Wigner perturbation theory.
The other $\lambda $ is for the case of $N_p = b$.
The value of this $\lambda $ is of 
 interest since
 topologically the structure of paths for the case of
 $N_p \ge b$ is different from that for $N_p <b$.
 For example, when $N_p \ge b$, paths
 starting from $j < p_1$ can never reach a $j'$ which is greater than
 $p_2$. That is to say, no path can cross the non-perturbative region. On the other hand, when $N_p < b$, 
paths can cross the non-perturbative region. 
One can expect that such a difference should have some effects on the shapes of both EFs and LDOS.

\section{Numerical study for the perturbative and non-perturbative
parts of eigenfunctions }

In this section we study numerically 
the division of EFs into perturbative (PT)
and non-perturbative (NPT) parts. Both the individual
shape and the average shape of EFs will be studied.

The boundary of the NPT part of a state $|\alpha \rangle $, 
i.e., the values of $p_1$ and $p_2$, are calculated in the 
following way. 
For a Hamiltonian matrix as in Eq.~(\ref{Hij}),
we first diagonalize it numerically and get all its EFs. 
Then, for an eigenstate $|\alpha \rangle $ with energy $E_{\alpha }$, we find out all the  
pairs of $p_1$ and $p_2$ for which 
the value of $\langle \alpha |(T^{\dagger})^n
T^n|\alpha \rangle $ could become smaller than a small
quantity (say, $10^{-6}$) when $n$ goes to infinity.
Finally, for the pair of ($p_1
, p_2$)  thus obtained with the smallest value of $N_p=p_2-p_1+1$ and the pairs of ($p_1+1,p_2$) and ($p_1,p_2-1$),
we calculate eigenvalues of the corresponding $U$ operators
in Eq.~(\ref{U}) in order to check out if 
all the $|\lambda u_{\nu}|$ for the pair of $(p_1,p_2)$ are 
less than 1 while 
some of the $|\lambda u_{\nu}|$ for each of the other two
pairs  $(p_1+1,p_2)$ and $(p_1,p_2-1)$ are larger than 1.

    The shape of an eigenstate $|\alpha \rangle $
 in the unperturbed  states can be defined as  
\begin{equation} 
W_{\alpha}(E^0 ) = \sum_{k} |C_{\alpha k}|^2 \delta (E^0-E^0_k). 
\label{ef} \end{equation}
In order to obtain the average shape of eigenstates, 
different individual distributions $W_{\alpha }(E^0)$ 
should be first shifted 
into a common region. 
 For this, we express $W_{\alpha}(E^0)$ with respect to 
$E_{\alpha}$  before averaging. For $\alpha $ from $\alpha _1$
to $\alpha _2$, with $W_{\alpha }(E^0)$ expressed as
$W_{\alpha }(E^0-E_{\alpha })$, the average shape of the 
eigenstates, denoted by $W(E^0_s)$, can be calculated.
Numerically, $W(E^0_s)$ are thus calculated: For a state
$|\alpha \rangle $, some value of $E^0_{s}$
and all $k$ satisfying $|E^0_{s} -(E^0_k-E_{\alpha })| < \delta E_w /2$,
where $\delta E_w$ is some quantity chosen for dividing $E^0_s$
into small regions,  we calculate the sum of 
$ \sum_k |C_{\alpha k}|^2$, then take the average of the sum over
$\alpha $ from $\alpha _1$ to $\alpha _2$. The result thus obtained 
is the average value of $W$ in the region $(E^0_{s}-\delta E_w/2,
E^0_{s}+\delta E_w/2)$ and is also denoted by $W(E^0_{s})$.
Similarly, the values of $W(E^0_s \pm m\delta E_w)$
($m=1,2, \cdots $) can also be calculated.
The average shape of eigenstates can also be divided into 
a NPT part and a PT part by the averaged boundary of 
the NPT parts of the states $|\alpha \rangle $, denoted by
$p_1^a \equiv \langle p_1-E_{\alpha } \rangle $ and 
$p_2^a \equiv \langle p_2-E_{\alpha }\rangle $. The 
size of the NPT part of the average shape of eigenstates is  
$\langle N_p \rangle  \equiv \langle p_2-p_1+1 \rangle $.  
The values of $\lambda $ for $\langle
N_p \rangle $ beginning to be larger than 1 and for $\langle
N_p \rangle =b$ will  be denoted by {\it $\lambda _f$ and
$\lambda _b$ }, respectively. 

Now let us present the numerical results. The first one is  for the case of
$N=300, b=10$ and $\lambda =0.6$. This is a case for which $N_p$ can
be equal to both 1 and 2. The values of
$|C_{\alpha k}|^2=|\langle k|\alpha \rangle |^2$
for four EFs in the middle energy region
($\alpha  = 148 - 151$)
are given in Fig.~1. The two vertical
dashed-dot straight lines for each case
indicate the positions of $p_1$ and $p_2$
of the NPT part of the  eigenstate, i.e.,
the boundary of the NPT part. Since $E^0_k=k$, $p_1$ and $p_2$
are also the unperturbed eigenenergies of the corresponding basis states
$|k=p_1 \rangle $ and $|k=p_2 \rangle $. For $\alpha =148$
and 149, since $p_1=p_2=\alpha $ there is only one vertical dashed-dot
line in each case. 
For $\alpha =150$ and 151, 
there are, in fact, two NPT parts for each case. 
 In Fig.~1 we give one of them for each
case only, i.e., $(p_1,p_2)=(149,150)$ for $\alpha =150$ 
(the other one is (150,151)) and 
$(p_1,p_2)=(151,152)$ for $\alpha =151$ (the other one 
is (150,151)). Numerically we have found that only for small
$\lambda $ is it possible for an eigenstate to have
more than one NPT parts. 
The average shape of EFs for $\alpha $ from 130 to 170 is given
in Fig.~2 with the corresponding values of $ p_1^a $
and $p_2^a $ indicated by vertical dashed-dot lines. $\delta
E_w = 1$ for calculating this $W(E^0_s)$ function. 

As mentioned in section II, another perturbation strength of interest
is for the case of $N_p $ being able to equal  the band
width $b$ of the Hamiltonian matrix. An example 
for this case is  $\lambda =1.4$.
For this value of $\lambda $, there is only one NPT part for each
state $|\alpha \rangle $, e.g., $N_p=9,10,7,9$ for $\alpha
=148,149,150,151$, respectively. 
Individual EFs for $\alpha =148 - 151$  are presented
in Fig.~3 with their boundaries of the NPT parts. Although for an
individual EF the value of $|C_{\alpha k}|^2$ could be still large
outside  the NPT region (Fig.~3),
the main body of the average shape of  EFs lies obviously
in the averaged NPT region as shown in Fig.~4(a).
For this averaged EF we still have $\delta E_w$ equal to 1 and
$\alpha $ from 130 to 170. An interesting feature
of the distribution $lnW$ in the slope regions $[p_1^a-b,p_1^a)$ and 
$(p_2^a,p_2^a+b]$ in Fig.~4(b)
is that, as predicted in section II,
the decaying speed in the two slope regions 
is obviously slower than that
in the long tail regions. 
From Fig.~4(a) it is quite clear why we call the regions
``slope''.

In Ref. \cite{CCGI96} the average shape of EFs for  
different values of the Wigner parameter $q$,
\be q=\frac{(\rho v)^2}{b}, \ee
and an ergodicity parameter ``$\lambda _e$'' (denoted by 
$\lambda $ in that paper),
\be \lambda _e = \frac{ab^{3/2}}{4\sqrt{2} c\rho v}
\ \ with \ a \approx 1.2, \  c \approx 0.92, \ee
 has been studied numerically.
Results in that paper show that when $q$ is small and $\lambda _e$ not small, e.g., $q= 1$ and
$\lambda _e= 3.7$,  two averaging procedures, namely, averaging
with respect to centers of energy shell, which are close to the
perturbed eigenenergies, and averaging with respect to centroids
of EFs, 
give similar results for the average shape of EFs; while
for large $q$ and small $\lambda _e$, e.g., $q=90$ and 
$\lambda _e=0.24$, the two averaging procedures give different results
and a so-called localization in  energy shell has been
found. 

What is of interest here is to study the difference 
between the above two cases of different parameters $q$ and 
$\lambda _e$ from the viewpoints
of the GBWPT. In Fig.~5, four examples of individual EFs are given
for $\lambda=4.0$, for which $q=1.6$ and $\lambda _e=1.82$.
When edge effects can be neglected, from the condition for Eq.~(\ref{TnTn}) 
to hold given in the above section, it can be shown that 
the eigenenergies $E_{\alpha }$ should be close to the 
centers of the corresponding NPT regions, 
that is, 
\be \label{ENp}  E_{\alpha } \approx 
\frac{p_2 -p_1 }2. \ee  
Here by the NPT region of an EF we mean 
the region of $E^0_k$ between the boundaries (dashed-dot lines) of 
the NPT part of the EF as shown in Fig.~5. 
Then, since as shown in Fig.~5
centroids of the EFs are generally not far from the  centers of the 
NPT regions, they are generally not far from 
the corresponding eigenenergies.
Therefore, in this case the averaging with respect to the centroids of 
the EFs and the averaging with
respect to the eigenenergies do not have
much difference, and the average shape of EFs obtained by the two 
methods should be more or less similar. However, since there are still 
some differences between the  centroids of the EFs and the 
eigenenergies, results of the two averaging methods 
can not be quite close to each other. In fact, 
as shown in Ref. \cite{CCGI96}, 
results of the method of averaging with respect to  
centroids of EFs should be  a little narrower than those of the other method. 

The average shape of EFs for $\alpha $ from 130 to 170 
and $\lambda =4.0$
is given in Fig.~6 with $\delta E_w =3.0$.  
(In this paper numerically we study the averaging
with respect to eigenenergies only.)
For this averaged EF the slope regions are even clearer.
The sizes of the two slope regions in Fig.~6 are  larger 
than b, the predicted one for a single EF, since the values of $N_p$ for
the NPT parts of the eigenstates $|\alpha \rangle $ taken for averaging
are not the same. In fact, the variance of the $N_p$ for the states
is about 1.56.

For studying the case of localization in energy shell as in Ref. \cite{CCGI96},
the dimension $N=300$ is too small. Therefore, we increase 
the value of $N$ to 900 and study the case of
$b=10$ and $\lambda =30.0$, for which the parameters
$q$ and $\lambda _e$ are equal to 90 and 0.24, respectively, as those in Ref. \cite{CCGI96}. 
The values of $|C_{\alpha k}|^2$ (dots) of four EFs
in the middle energy region are given in Fig.~7 with the
corresponding boundaries ($p_1$ and $p_2$) of the NPT parts of
the EFs (indicated by  vertical dashed-dot lines). Clearly
the main body of each of the EFs occupies only part
of the NPT region. Then, resorting to Eq.~(\ref{ENp}), it is easy to see
that for the average shape of EFs the result of averaging with respect to
centroids of EFs should be obviously narrower than that of 
averaging with respect to eigenenergies. 
This phenomenon is called localization in energy shell 
in Ref. \cite{CCGI96}. 
In our opinion, it is, in fact,  
localization of EFs in their NPT regions. 

Although, as shown in Fig.~7, when $\lambda =30.0$ 
many components of the EFs are quite small in the NPT parts, 
 they can still be distinguished from those in the PT parts. 
In Fig.~8 we present the values of $|C_{\alpha k}|^2$ 
in Fig.~7 in logarithm scale, from which the difference 
between the $|C_{\alpha k}|^2$ in the NPT parts and those in 
the PT parts is quite clear. In fact, it can be seen from Fig.~8 that 
$|C_{\alpha k}|^2$ for small components in the NPT parts also
decay exponentially on average, but the decaying speed 
is slower than that for components in the PT parts. 
 Similar results have also 
been found for other eigenfunctions and the centroids 
of the EFs have been found scattered 
in the NPT regions.
Therefore, averaging with respect to eigenenergies
(around the centers of the NPT regions) is reasonable when
studying the average shape of EFs.
The average shape of EFs in the middle energy region
for $\alpha $ from 420 to 480 is shown in Fig.~9 ($\delta E_w
=10$). The two slope regions are also obvious. 
The widths of the two slope regions, which can be
seen from the figure is about $2\delta E_w =2b$,  are also larger
than that predicted for a single EF 
due to the fact that the
variance of $N_p$ for these states $|\alpha \rangle $ is about 5.14.

In order to have a clear picture for the variation of the 
average size $\langle N_p \rangle $ 
of NPT parts of EFs with the perturbation strength $\lambda $, we plot it in Fig.~10(a) 
(for $N=300$ and $\alpha $ from
130 to 170).
The two solid straight lines in the figure show that $\langle N_p
\rangle $ has a good linear dependence on $\lambda $ in the two regions
of $\lambda \in (2.0,4.5)$ and $\lambda \in (4.5,7.0)$.
The variation of $\langle N_p \rangle $ for $\lambda $ 
less than 2 is given in 
the inset of the figure in more detail. 
The solid curve in the inset is a fitting curve of 
a quadratic form: ($7.5(\lambda -0.4)^2+1.0$). 
Since in this case the value of $\lambda _f$ is a little larger than 0.4, i.e., for $\lambda \le 0.4 $ 
the size $N_p$ of the NPT parts of the EFs is equal to one, 
the region of $\lambda < 0.4$ has been neglected  
when fitting the $\langle N_p \rangle $ by the quadratic curve. 
From the inset it can be
seen that the fitting curve fits the values of $\langle N_p \rangle $ 
quite well in the region of $\lambda \in (0.4,1.5)$. 
Interestingly, the value of $\lambda _b$
 for $\langle N_p \rangle =b$ is between 1.45 and 1.5, 
that is to say, in the region of $\lambda \in (\lambda_f 
, \lambda _b)$, the values of $\langle N_p \rangle $ 
have a quadratic dependence on $\lambda $. 

The variance of $N_p$, denoted by $\Delta N_p$, 
 is given in Fig.~10(b) (circles), which  shows that 
$\Delta N_p$ is small compared with $N_p$. From the figure it can be seen 
that for $\lambda < 5$ the value of $\Delta N_p$ increases
with $\lambda $ on average. Then, 
since, as mentioned above, the value of $\Delta N_p $ is 
about 5.14 when $\lambda =30$, it does not change much in the region
of $\lambda \in (5,30)$, 
that is to say, the relative value of the variance, namely
$\Delta N_p /N_p$, becomes even smaller as $\lambda $ becomes larger. 

\section{Relationship between the shape of eigenfunctions and of LDOS}

The existence of a relationship between the average shape
of EFs and that of LDOS in some particular cases 
has already been noticed numerically 
by several authors (see, for example, \cite{CCGI96,BGIC97,WIC98}).
Here for the WBRM we show that some analytical 
arguments can be given for it by making use of the GBWPT. 
                                                        
    The so-called  local spectral density of states (LDOS) for an 
unperturbed state $|k\rangle $  is defined as 
\begin{equation} 
\rho _L^k(E) = \sum_{\alpha} |C_{\alpha k}|^2 \delta (E-E_{\alpha})
\label{ldos} \end{equation}
where   $C_{\alpha k} = \langle k|\alpha \rangle $.
Making use of the division of EFs into NPT and PT parts, 
the non-perturbative (NPT) and perturbative (PT) part 
of the LDOS can be defined in the following way.   
Let us consider all the perturbed states $|\alpha \rangle $ 
for which $C_{\alpha k} $ is in the NPT parts of their eigenfunctions. 
Suppose $\alpha _{p_1} $ is the smallest one 
among these  $\alpha $  and $\alpha _{p_2}$ 
is the largest one among them, i.e., $|\alpha _{p_1} \rangle $ 
has the  smallest eigenenergy and 
 $|\alpha _{p_2} \rangle $ has the largest.
A property of the WBRM is that for any state 
$|\alpha \rangle $ with $\alpha _{p_1} \le \alpha \le \alpha _{p_2}$, 
generally to say, $C_{\alpha k}$ is in the NPT part of 
its eigenfunction.  
Then, the non-perturbative (NPT) part 
of the LDOS $\rho _L^k(E)$ can be defined as the sum 
in Eq.~(\ref{ldos}) over $\alpha \in [\alpha _{p_1}, 
\alpha _{p_2}]$, and 
the perturbative (PT) part can be defined as the sum 
over the rest $\alpha $. 
The size of the NPT part of the LDOS
is defined as $(\alpha_{p_2} -\alpha _{p_1}+1)$ and  
will be denoted by $N_{\alpha _p}$.

The average shape of LDOS, denoted by $\rho _L(E_s)$,
 can be obtained in a way
similar to that for the EFs discussed in 
section III, except that for the LDOS the distributions
$\rho_L^k(E)$ are expressed
as functions of $(E-E^0_k)$ before averaging. 
Since $E^0_k$ is also the centroid of $\rho _L^k(E)$,
such an averaging method is the most natural one.
The average shape of LDOS can also be divided into a NPT 
part and a PT part by the averaged boundary of the 
NPT parts of the individual LDOS. 

Before discussing the relation between the average 
shape of EFs and that of LDOS, let us first draw 
some conclusions for properties of EFs from the numerical 
results discussed in section III. Firstly, 
based on the numerical results given in section III
for the average shape of EFs,
we can make such an approximate treatment for 
the average values of $|C_{\alpha i}|^2$
{in the NPT parts of EFs}: for a given $\lambda $ not 
extremely large, they can be treated as
a constant denoted by $t_{\lambda }^2$, i.e., $\overline  
{|C_{\alpha i}|^2 } \approx t_{\lambda }^2$,
  when edge effects can be neglected.
Here we would like to stress that Figs.~2,4,6,8 show that 
the values of $\overline {|C_{\alpha i}|^2}$ on the edges 
of the NPT parts become smaller than those in the middle 
regions when $\lambda $ increases and 
the above approximate treatment may fail in case of 
extremely large $\lambda $. Secondly, another result of the last
section is that the variance $\Delta N_p$
is small compared with $N_p$. 
This means that when discussing
approximate relations 
it is reasonable to treat the size
$N_p$ of NPT parts of eigenstates as a constant
for a fixed  $\lambda $. 

As results of the above two approximations 
and the approximate relation (\ref{ENp}), it can be shown 
that, under the corresponding conditions, 
(a) the average value of $|C_{\alpha k}|^2$ in the NPT parts of LDOS is also about $t_{\lambda }^2$, 
 (b) the size of the NPT part of a LDOS is close to that 
of an EF, i.e., $N_{\alpha _p} \approx N_p$, (c) usually 
$E_k^0$ is in the middle region 
of the NPT part of the LDOS $\rho_L^k(E)$, i.e., $E^0_k - 
E_{\alpha _{p1}} \approx E_{\alpha _{p2}} -E^0_k $.
Then, one can see that 
the following approximate relation holds
for the NPT part of an averaged EF $W$ and the NPT part of an 
averaged LDOS $\rho _L$, 
\be \frac{W(E)}{\rho ^0(E)} \approx \frac{\rho_L(E)}{ \rho(E)}
\label{EFLD1} \ee
where $\rho ^0(E)$ and $\rho (E)$ are the averaged 
density of states of the
unperturbed and perturbed spectra in the corresponding regions,
respectively,. Here, for brevity, instead of $E^0_s$ and $E_s$, we use $E$
to denote the variables of the functions $W$ and $\rho _L$. 

Next let us discuss the relation between the PT part of 
the average shape of EFs and the PT part of the average shape of LDOS. 
According to Eq.~(\ref{Cf}), $C_{\alpha j}^2$ for a basis state
$|j \rangle $ in the $Q_{\alpha }$ subspace 
can be written as 
\begin{equation}
C_{\alpha j}^2 = \sum_{i_1=p_1}^{p_2} 
\sum_{i_2=p_1}^{p_2}
\sum_{s_1,s_2}
f_{\alpha s_1}(j \to i_1) f_{\alpha s_2}(j \to i_2)
C_{\alpha i_1}C_{\alpha i_2}.
\label{Cf2} \end{equation}
Following a way similar to that of calculating the average shape 
of EFs $W(E_s)$ in the last section, we take the average of this quantity over different eigenfunctions. 
Since $C_{\alpha i}$ in the 
NPT parts of EFs have random signs, the main
contribution to $\overline {C_{\alpha j}^2}$ in 
Eq.~(\ref{Cf2}) comes from the
$i_1=i_2$ terms and 
\begin{equation} \label{ac2} 
\overline {C^2_{\alpha j}}  \approx \sum_{i=p_{ 1}}^{p_{ 2}} 
 \sum_{s_1,s_2} \overline {f_{\alpha s_1} (j \to i)
 f_{\alpha s_2} (j \to i)
} t^2_{\lambda }. 
\end{equation} 
When the number of eigenstates taken for averaging is large enough,
since $f_{\alpha s}$ is a product of factors $V_{kk'}/(E_{\alpha }
-E^0_k)$, the main contribution in (\ref{ac2}) to $\overline
{C_{\alpha j}^2}$ comes from the terms with path $s_1$ equal to
path $s_2$, that is, 
\begin{equation} \label{ac2-1} 
\overline {C^2_{\alpha j}}  \approx  \sum_{i=p_{ 1}}^{p_{ 2}} 
 \sum_{s} \overline {f_{\alpha s}^2 (j \to i)
} t^2_{\lambda }. 
\end{equation}

Let us first discuss the relation (\ref{ac2}), which 
is for cases without enough eigenstates taken for averaging. In these cases, 
due to the signs of the denominators
$(E_{\alpha } -E^0_k)$ of the factors of paces, 
there is a systematic difference between the contribution of paths 
starting from $j$ with
$E^0_j < E_{\alpha }$ and the contribution of paths starting from
$j'$ with $E^0_{j'} > E_{\alpha }$, especially when $N_p \ge b$.
A result of this difference is that $\overline {|C_{\alpha j}|^2}$ for  
the left PT part of an averaged EF may be different 
from the corresponding one for its right PT part. 
But, for PT parts of averaged EFs on the same side,
such a difference does not exist. 
In fact, when edge effects can be neglected, 
the structure of paths is similar for different EFs 
when the difference in the size of NPT parts of the EFs can 
be neglected, 
and we have
\begin{equation} \label{ff} 
\overline{ f_{\alpha s_1}(j \to i) f_{\alpha s_2} (j \to i)} \approx 
\overline{f_{\alpha ' s_1'} (j' \to i') f_{\alpha ' s_2'} (j' \to i')} 
 \end{equation} 
where $s_1'$ $s_2'$ are paths similar to $s_1$ and $s_2$, 
respectively, but from $j'$ to $i'$ 
with $j'-i'=j-i$ and $j'-p_1'=j-p_1$. 
From the relations (\ref{ac2}) and (\ref{ff}), one can see
that, when edge effects can be neglected, the left (right) PT parts 
of averaged EFs in different energy regions should be similar. 
By definition, the left (right) PT part of the average
shape of EFs 
should be related to the right (left) PT part of the average
shape of LDOS, since, for example, a $|C_{\alpha j}|^2$ in the
left PT part of the EF $W_{\alpha }(E^0) $ is in the 
right PT part of the LDOS $\rho^j_L(E) $. Then, 
the relationship between the PT part of the average shape 
of EFs and that of the average shape of  LDOS should be 
\be \frac{W(E)}{ \rho^0 (E)} \approx \frac{\rho_{-L}(E)}{ \rho(E)}
\label{EFLD2} \ee
where $\rho _{-L}(E)$, the inverted LDOS, is defined as $\rho _{-L}(E)
\equiv \rho_L(-E)$. This means that the left (right) 
tail of the average shape of EFs is more similar to the right
(left) tail of the average shape of LDOS than its right
(left) tail does. Since $\overline {|C_{\alpha i}|^2}$ 
can be approximately treated as a constant for the NPT parts 
of both EFs and LDOS, the approximate relation 
(\ref{EFLD2}) also 
holds for the NPT parts of the average shape of EFs and of LDOS. 

When there are enough eigenstates taken for averaging, we have
the relation (\ref{ac2-1}). In this case, instead of (\ref{ff}), we have
\begin{equation} \label{ff-1} 
\overline{ f_{\alpha s}^2 (j \to i) } \approx 
\overline{f_{\alpha ' s'}^2 (j' \to i') },
 \end{equation}
and  the left PT part of an averaged EF should be 
similar to its right PT part. 
Then, the same relation as in (\ref{EFLD1}) can
be obtained for the PT part of the average shape of EFs and the
PT part of the average shape of LDOS. In this case, $\rho _L(E_s) \approx
\rho _{-L}(E_s)$.

The above results have been checked by numerical calculations.
In Fig.~11(a) and (b), 
an average shape of EFs $W(E)$ (circles) is compared 
with a corresponding inverted average shape of  LDOS $\rho _{-L}(E)$ 
(solid curve) for $\lambda =4.0$ and $b=10$.
In order to avoid edge effects, $N$ is chosen to be 900. 
In this case, in the middle energy regions the difference 
between $\rho (E)$ and $\rho ^0(E)$ can be neglected.  
The averaging
has been done for 60 perturbed and 60 unperturbed states in the
middle energy regions, respectively. 
For this relatively small
number of states taken for averaging,
 one can expect that the relation (\ref{EFLD1})
is not so good as the (\ref{EFLD2}) in the tail region. 
Indeed, as shown in
Fig.~11(c) the LDOS $\rho _L(E)$ (solid curve)
is not so close to the $W(E)$ (circles) as the inverted one
$\rho _{-L}(E)$  in Fig.~11(b). In order to show the influence of edge
effects on the relation (\ref{EFLD2}), in Fig.~11(d) we give
a result obtained when $N$ is equal to 300. 

Finally, we would like to stress that although the above 
arguments leading to the approximate relations 
(\ref{EFLD1}) and (\ref{EFLD2}) 
are for not very large $\lambda$, 
numerical results in Ref. \cite{CCGI96} show that  
central parts of averaged EFs (for averaging with respect 
to eigenenergies) are still close to those of averaged LDOS
when $\lambda $ is quite large. 
In fact, for a given $\lambda $,
the approximation $\overline 
{|C_{\alpha i}|^2} \approx t_{\lambda }^2$ for NPT parts
of EFs, which has been employed in the above arguments, 
is not necessary for deducing the relations
(\ref{EFLD1}) and (\ref{EFLD2}). For example, under the  approximation that 
$\overline {|C_{\alpha i_1}|^2} \approx 
\overline {|C_{\alpha i_2}|^2}$ with $i_1-p_1=p_2-i_2$
for NPT parts of EFs, 
similar arguments as given above also lead to the two 
relations (\ref{EFLD1}) and (\ref{EFLD2}). In fact, this approximation
is in even better agreement with the numerical results 
given in the last section than the former approximation
$\overline {|C_{\alpha i}|^2} \approx t_{\lambda }^2$
and may be still valid for very large $\lambda $. 
Therefore, the two relations (\ref{EFLD1}) and (\ref{EFLD2}) 
may be correct for very large $\lambda $, too.

\section{Variation of the shape of LDOS with perturbation strength}

Analytical and numerical study for the shape of the LDOS of the
WBRM has been done in, e.g., \cite{Wigner,FGGK94,FCIC96,CFI}. It
is known that when perturbation is weak (but not very weak) the average 
shape of LDOS is of the Breit-Wigner form (Lorentzian distribution),
and when perturbation becomes strong the shape will approach to
a form predicted by the semi-circle law.

Our interest here is in studying the process of the transition from the
Breit-Wigner form to the semi-circle
 form, in order to see if the GBWPT can throw light on how 
the transition occurs.
In fact, due to two results of the previous two
sections that (a) the central part of the average shape of
EFs is composed of its NPT part and the slope region of the PT part 
and (b) the average shape of EFs is similar to that of LDOS when
the density of states of the perturbed spectrum is similar to
that of the unperturbed spectrum, 
one can expect that properties of the 
average shape of LDOS, particularly of its half-width, 
should be related to properties of 
 $\langle N_p \rangle $, 
the average size of NPT parts of EFs. 

An interesting property of the semicircle law
\be \label{SC} \rho _{sc} (E) = \frac 2{\pi R^2_0} \sqrt {R^2_0 -
E^2}, \ \ \ \ \ \ \ \ |E| \le R_0, \ee
where $R_0=\lambda \sqrt{8b}$, is that it obeys a scaling law. Specifically,
after a rescaling
\be  R_0 = \lambda R_0', \ E=\lambda E', \ 
\rho_{sc}(E) = \rho_{sc}'(E')/\lambda ,  \ee
it becomes
\be \label{SC1} \rho _{sc}' (E') =
\frac 2{\pi {R_0'}^2} \sqrt {{R_0'}^2 -
{E'}^2}, \ee
having the same form as the $\rho_{sc}$ in Eq.~(\ref{SC}). In fact, 
Eq.~(\ref{SC1}) is just Eq.~(\ref{SC}) for the case of $\lambda =1$.
This property of the semicircle law supplies a convenient method of
studying the approaching of the average shape of LDOS $\rho_L(E_s)$
(for brevity, as in the last section,
the subscript $s$ for the variable $E_s$ will be omitted in what follows)
to the semicircle form $\rho _{sc}(E)$ with increasing  $\lambda $.
That is, first we change the LDOS $\rho_L(E)$ for a 
perturbation strength $\lambda $
to a rescaled one $\rho_L^{rs}(E)$ defined by
\be \label{ldos-rs} \rho_L^{rs}(E)=\lambda \rho_L(\lambda E),
\ee
then, we compare it with the semicircle form $\rho_{sc}(E)$
for $\lambda =1$. 

In Fig.~12 we present such  comparisons for
$\lambda $ from 1.2 to 1.9. The LDOS $\rho _L(E)$ in this and the 
following figures are for 
$N=500$ and $b=10$. 
In order to take average, 30 Hamiltonian matrices of different
realizations of the random numbers have been diagonalized for
each $\lambda $ and 50 individual LDOS $\rho _L^k(E)$ in the middle energy
region have been taken for each Hamiltonian matrix. 
Figure 12 shows that the sign for the
$\rho_L^{rs}(E)$ to approach the semicircle form appears
when $\lambda $ is between 1.5 and 1.6. 
Interestingly, in this case the value of $\lambda _b$ for $\langle N_p 
\rangle = b$ is also between 1.5 and 1.6. 
As indicated in section II, such a value of $\lambda $ 
is of interest, since for a perturbed eigenstate 
with $N_p  \ge b $ paths starting from the left
PT part of the state can not reach the right PT part, and vice versa.
That is to say, 
when $\lambda $ becomes larger then $\lambda _b$,  
there will be a  change in the topological structure of the paths. 
Therefore, such a coincidence 
should not be  an accident. 

When $\lambda $ becomes larger,
as is known, the form of $\rho_L^{rs}(E)$ will become 
closer to the semicircle form (Fig.~13). 
Since the semicircle law has a scaling behavior,
an interesting result of the comparisons between the rescaled LDOS $\rho_L^{rs}(E)$ and the semicircle form 
given in Figs.~12 and 13 is that 
the LDOS $\rho_L(E)$ obeys an approximate scaling law 
when $\lambda $ is larger than $\lambda _b$.  
As a consequence, the dependence of the half-width 
of the LDOS on $\lambda $ should become linear when 
$\lambda $ exceeds $\lambda _b$. 

The Breit-Wigner form
\be \label{BW} \rho_{BW}(E) = \frac{\Gamma / 2\pi}
{E^2+\Gamma ^2/4} \ee
does not have the scaling property as the semicircle law.
For the purpose of studying the relationship between the Breit-Wigner form
and the average shape of LDOS, we use the former as a 
fitting curve for the later with the width $\Gamma $ 
as the fitting parameter. The fitting
is done by requiring the minimum of the area difference
\be \Delta S = \int |\rho_L(E)- \rho_{BW}(E)| dE
\label{ds} \ee
between the Breit-Wigner form and the histogram of the averaged LDOS.
In numerical calculations, we first change 
the Breit-Wigner form to a histogram corresponding to 
the histogram of the LDOS, then calculate the $\Delta S$. 

Four examples thus obtained  for 
$\lambda =0.4,0.7,1.0$ and 1.5 are given in Figs.~14 and 15.
From numerical results we have found that the average shape 
of LDOS begins to  be fitted well by the 
Breit-Wigner form just before $\lambda $ reaches 0.4.
Interestingly, the value of $\lambda _f$ for $\langle N_p \rangle $
beginning to be larger than 1 is a little larger than 0.4.
That is to say, the LDOS of the Breit-Wigner form  begins to appear
just before the
ordinary Brillouin-Wigner perturbation theory fails. 
With the increasing of $\lambda $, the closeness between 
the LDOS and the Breit-Wigner form maintains
for a small region of the $\lambda $.
Then, when $\lambda $ increases further,
deviation will become more obvious. In fact, 
when $\lambda = \lambda _b \approx 1.5$, the LDOS 
$\rho_L(E)$ is absolutely
different from the Breit-Wigner form, while it becomes closer
to the semicircle form.

Since the smallest
area difference $\Delta S$ in Eq.~(\ref{ds}) gives a
measure for the deviation of the average shape of LDOS from the
Breit-Wigner form, we plot it in Fig.~16 (squares). It 
can be seen that the deviation
reaches its saturated value at about $\lambda =2.0$.
Similarly, in order to show the deviation of the average shape of LDOS from
the semicircle form, one can introduce another area difference  
\be \Delta S = \int |\rho_L^{rs}(E)- \rho_{sc}(E)| dE
\label{ds2} \ee
where $\rho_{sc}(E)$ is for $\lambda =1$. Variation of this
$\Delta S$ with $\lambda $ is also given in Fig.~16 (circles).
It shows that before $\lambda $ reaches 1.5 the deviation is large
and drops quickly. For $\lambda$  from 1.5 to about 4.5, $\Delta S$
drops slower. When $\lambda $ is larger than  4.5, the values
of $\Delta S$ are quite small and change quite slowly, which means
that the average shape of LDOS has become quite close to the semicircle form. 

Variation of the half-width of the average shape of LDOS is of
both  experimental and theoretical interest. Such a variation
is given in Fig.~17 by triangles. As expected, when $\lambda $
exceeds $\lambda _b \approx 1.5$, the half-width
can be fitted  well by a straight line. 
The variation of $\langle N_p \rangle $ with $\lambda $ is also
given in Fig.~17 (circles). Corresponding to the three regions of
$\lambda $ for the variation of the $\Delta S$ measuring
the deviation of the LDOS from the semicircle form 
represented by circles in Fig.~16, the 
variation of $\langle N_p \rangle $ can also be divided into
three regions: (1) $\lambda < 1.5$,
(2) $1.5 < \lambda < 4.5$, in which it can be fitted well by a straight
line, and (3) $\lambda > 4.5$, in which it can be fitted well by another
straight line. 

The fitting for the half-width of the LDOS and for
the $\langle N_p \rangle $ for small
$\lambda $ is given in the upper-left inset in Fig.~17. Since when $\lambda
< 0.4$ the value of $\langle N_p \rangle $ equals to 1, the
fitting for $\langle N_p \rangle $ (circles) has been done for
$\lambda >0.4$ by
a quadratic curve $(a(\lambda -0.4)^2 -1.0)$ with $a 
\approx 7.0$. The quadratic
feature of the dependence of $\langle N_p \rangle $ on $\lambda $
is quite clear in the region of $0.4 < \lambda < 1.5$. For the
half-width (triangles), the same form of fitting curve with $a \approx
15.0 $ is also given in the inset. 
As expected, the quadratic feature is
also clear for the half-width in the region of $0.4 < \lambda < 1.5$. Therefore, 
the correspondence between behaviors of the average size
of NPT parts of EFs and those of the half width of the 
average shape of LDOS is quite clear. 
The lower-right inset in Fig.~17
gives a comparison between the $\langle N_p \rangle $
for $N=500$ (solid curve) here and the $\langle N_p \rangle $
 for $N=300$ (dashed line)
in Fig.~10. The small deviation between them comes from both
edge effects and the fact that they are for Hamiltonian matrices
with different values of off-diagonal elements.

In conclusion, as expected, numerical results given in this section
show that
properties of the average shape of LDOS are indeed related to
properties of the average size of NPT parts of perturbed eigenstates and
knowledge of the latter, especially the value of $\lambda _b$
for $\langle N_p \rangle =b$, can indeed give deeper
understanding for the former.

\section{Conclusions}

The Wigner Band Random Matrix (WBRM) model is studied numerically
in this paper by making use of a generalization of the
Brillouin-Wigner perturbation theory (GBWPT). According to the
GBWPT, an energy eigenfunction (EF) of a perturbed system
can be divided into a non-perturbative
(NPT) and a perturbative (PT) part with the PT part expressed as
a perturbation expansion. Further more, the PT part can be divided
into a slope region and a tail region.

Numerically we have found that
for the average shape of EFs
 its central part is composed of its
NPT part and the slope region of its PT part. That is, the GBWPT
can give an analytical definition for the division of 
the average shape of  EFs into
central parts and tails. For the shape of individual EFs,
numerical results show that when the perturbation is not strong
their NPT part and the slope region of their PT part are usually
composed of large components. But when the perturbation becomes stronger,
the region of the NPT part of an EF occupied by large components
will become relatively smaller, i.e., the ratio of the region to the
whole NPT part will become smaller. As for the small components
in the NPT part of an EF in this case, 
the difference between them and the 
components in the PT part is also obvious.
Here we would like to point out that the possibility of dividing EFs
into NPT and PT parts should be useful in reducing calculation
time for eigenfunctions in a given energy region. 

It is already known from some previous numerical studies that
there is a relationship between the average shape of EFs
and that of LDOS, but the reason is not clear. Resorting to the
GBWPT and some conjectures made from numerical results, it is
possible to show that such a relationship indeed exist for the WBRM. Particularly,
when the number of states taken for averaging is not large
enough, the relationship is between the average shape of EFs
and the average shape of inverted LDOS.

A result of the above properties of the average shape of EFs and
of LDOS
is that some properties of the average shape of LDOS
should be related to properties of the NPT part of the average shape of EFs.
This has been studied in detail by numerical calculations.
Firstly, it is found that the
LDOS of the Breit-Wigner form  appears just before the 
perturbation strength $\lambda $ reaches a value $\lambda _f$,
for which the NPT parts of EFs
begin to have more that one components. Secondly, when 
$\lambda $ reaches $\lambda _b$, for which 
the average size $\langle N_p \rangle $
of NPT parts of EFs is equal to the band width $b$ of the 
Hamiltonian matrix,  
the average shape of  LDOS begins to be close to the
semicircle form predicted by the semicircle law, 
and for $\lambda $ larger than $\lambda _b$ the average shape of 
LDOS obeys an approximate scaling law. Thirdly,
variation of the half-width of the average shape of LDOS with perturbation strength
is closely related to the variation of the average size
$\langle N_p \rangle $ of NPT parts of EFs.
Particularly, when $\lambda < \lambda _b$, i.e., $\langle N_p \rangle  < b$,
both of them are of quadratic form, 
 and when $\lambda $ is larger than $\lambda _b$, both of them
 becomes linear.

\begin{center} {\bf Acknowledgements} \end{center}

We are very grateful to G.Casati and F.M.Izrailev  
for valuable discussions. 
This work was partly supported by
the National Basic Research Project ``Nonlinear Science'', China, 
Natural Science Foundation of China, 
grants from the Hong
Kong Research Grant Council (RGC) and the Hong Kong Baptist 
University Faculty Research Grant (FRG).

\begin{figure}
\narrowtext
\caption{Circles show the values of $|C_{\alpha k}|^2=|\langle
k|\alpha \rangle |^2$
for four energy eigenstates $|\alpha \rangle $ 
when $\lambda =0.6$ and $N=300$. 
The vertical dashed-dot  lines indicate 
the boundaries $p_1$ and $p_2$ of the
corresponding four non-perturbative parts of the eigenstates
(for $\alpha =148$ and 149, $p_1=p_2=\alpha $). }
\end{figure}   

%Fig.2
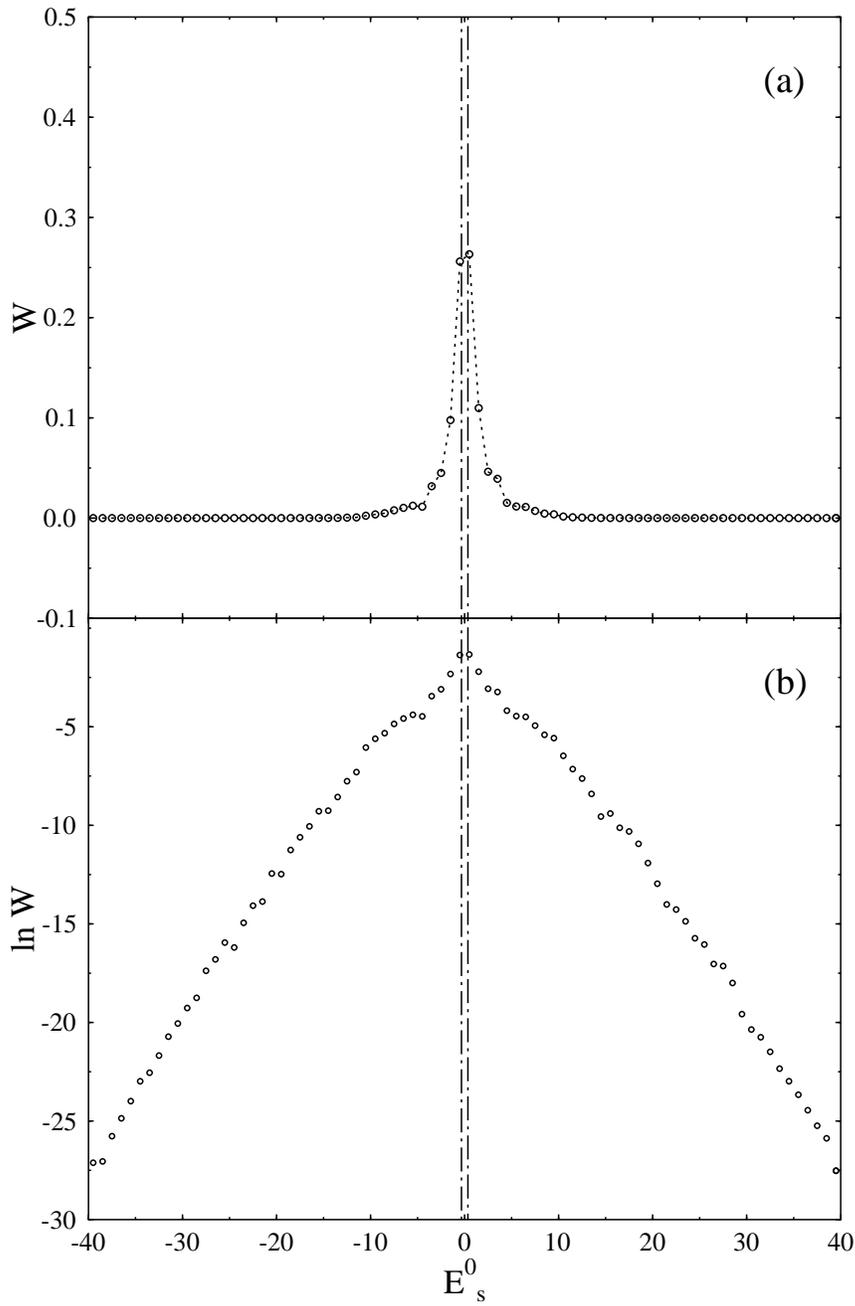
\begin{figure}
\narrowtext
\caption{ (a) The average shape of eigenstates in the middle energy
region, $W(E_s)$ (circles), for
$\lambda =0.6$ and $N=300$. 
The vertical dashed-dot lines indicate the averaged boundary
$p_1^a$ and $p_2^a  $ of the
corresponding non-perturbative parts of the eigenstates.
(b) Same as in (a) in  logarithm scale. }
\end{figure}   

%Fig.3
\begin{figure}
\narrowtext
\caption{Same as in Fig.~1 for $\lambda =1.4$. }
\end{figure}   

%Fig.4
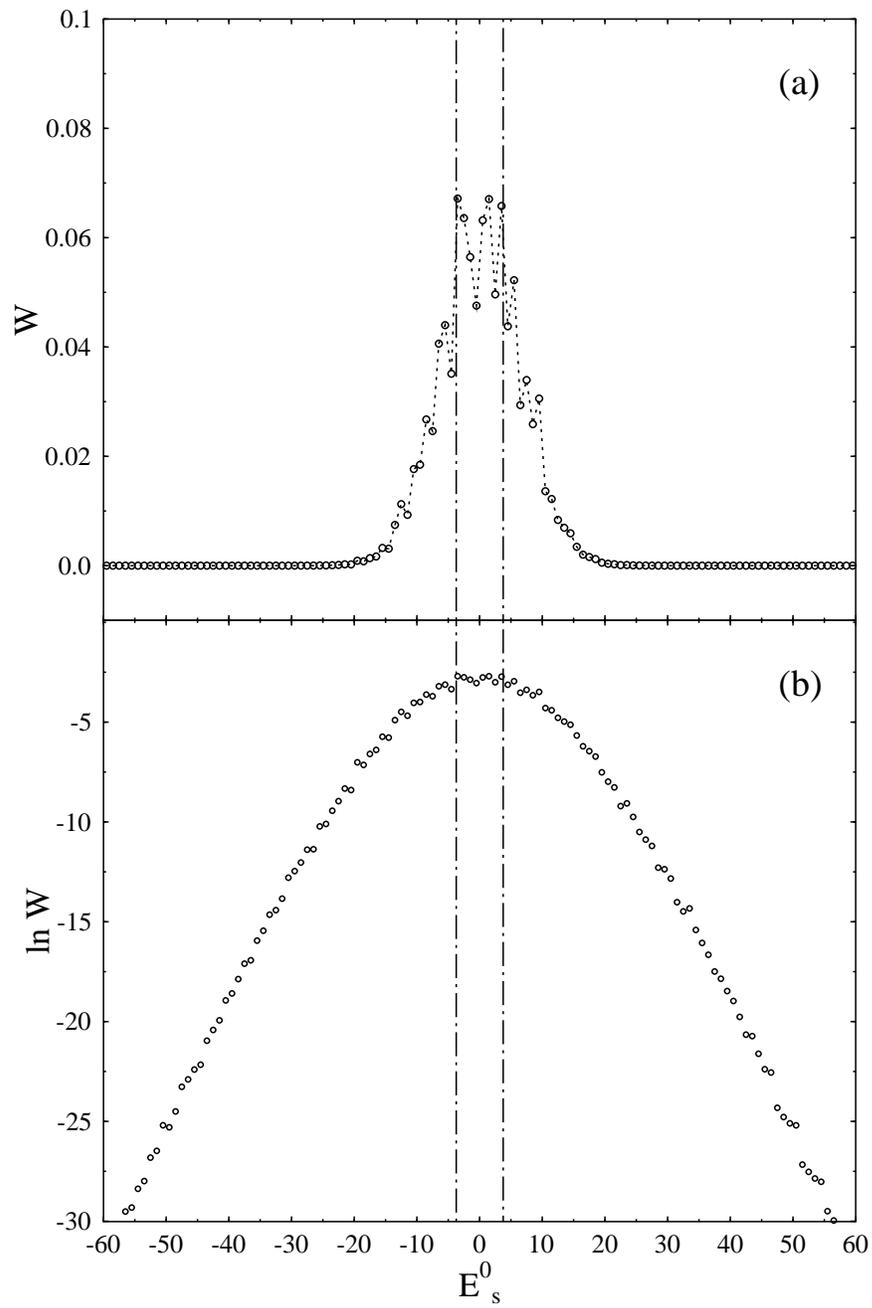
\begin{figure}
\narrowtext
\caption{Same as in Fig.~2 for $\lambda =1.4$. }
\end{figure}   

%Fig.5
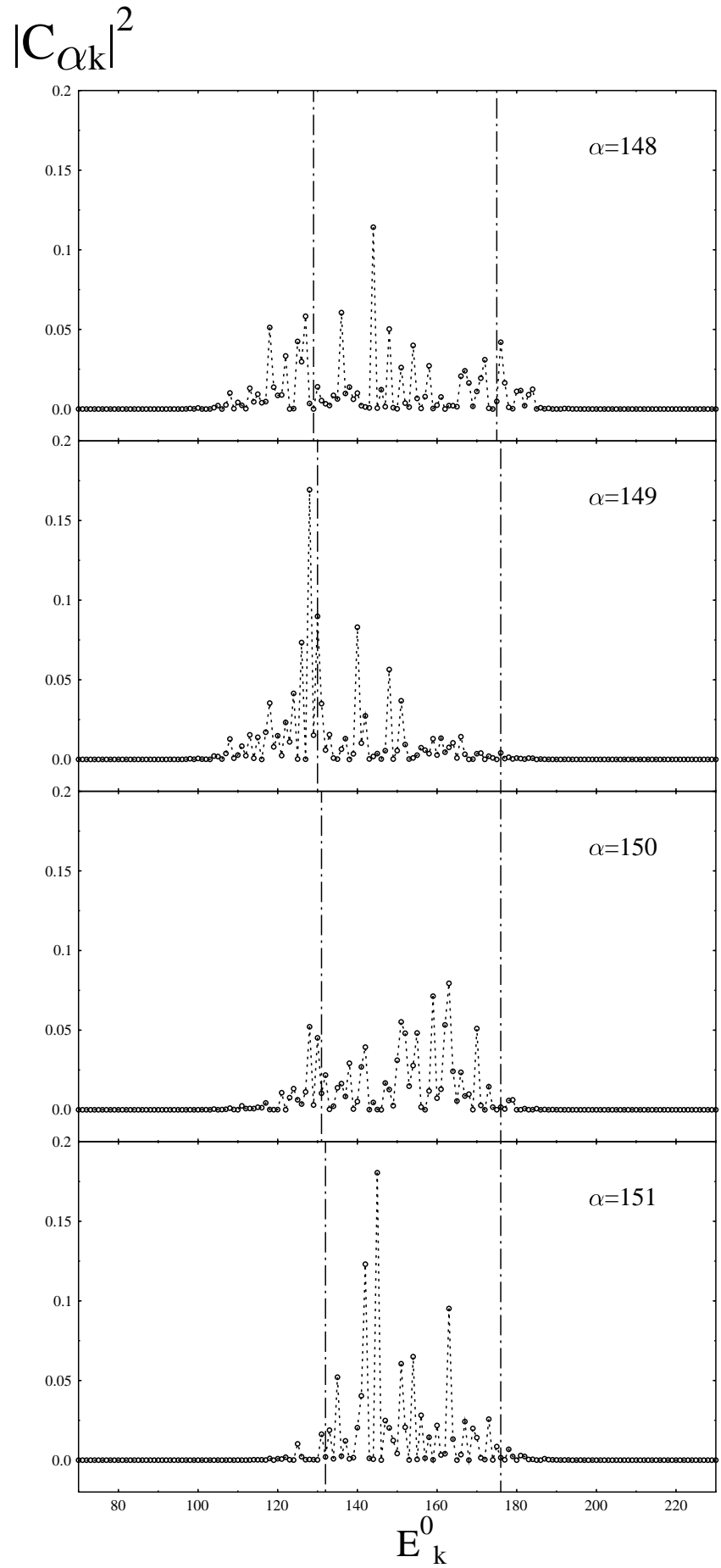
\begin{figure}
\narrowtext
\caption{Same as in Fig.~1 for $\lambda =4.0$. }
\end{figure}   

%Fig.6
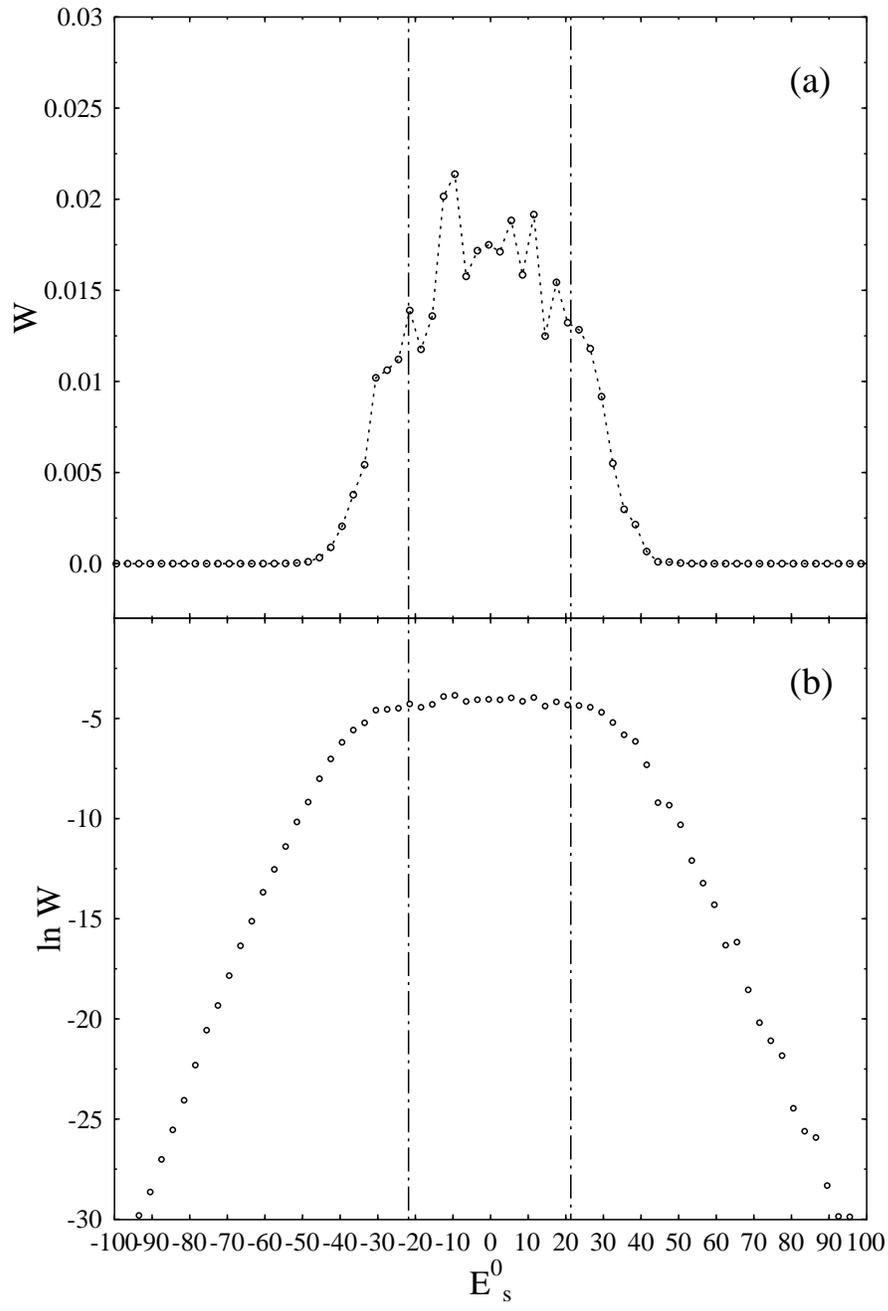
\begin{figure}
\caption{Same as in Fig.~2 for $\lambda =4.0$. }
\end{figure}

%Fig.7
\begin{figure}
\caption{Same as in Fig.~1 for $\lambda =30.0$ and $N=900$. }
\end{figure}   

%Fig.8
\begin{figure}
\caption{Same as in Fig.~7 in logarithm scale.  }
\end{figure}

%Fig.9
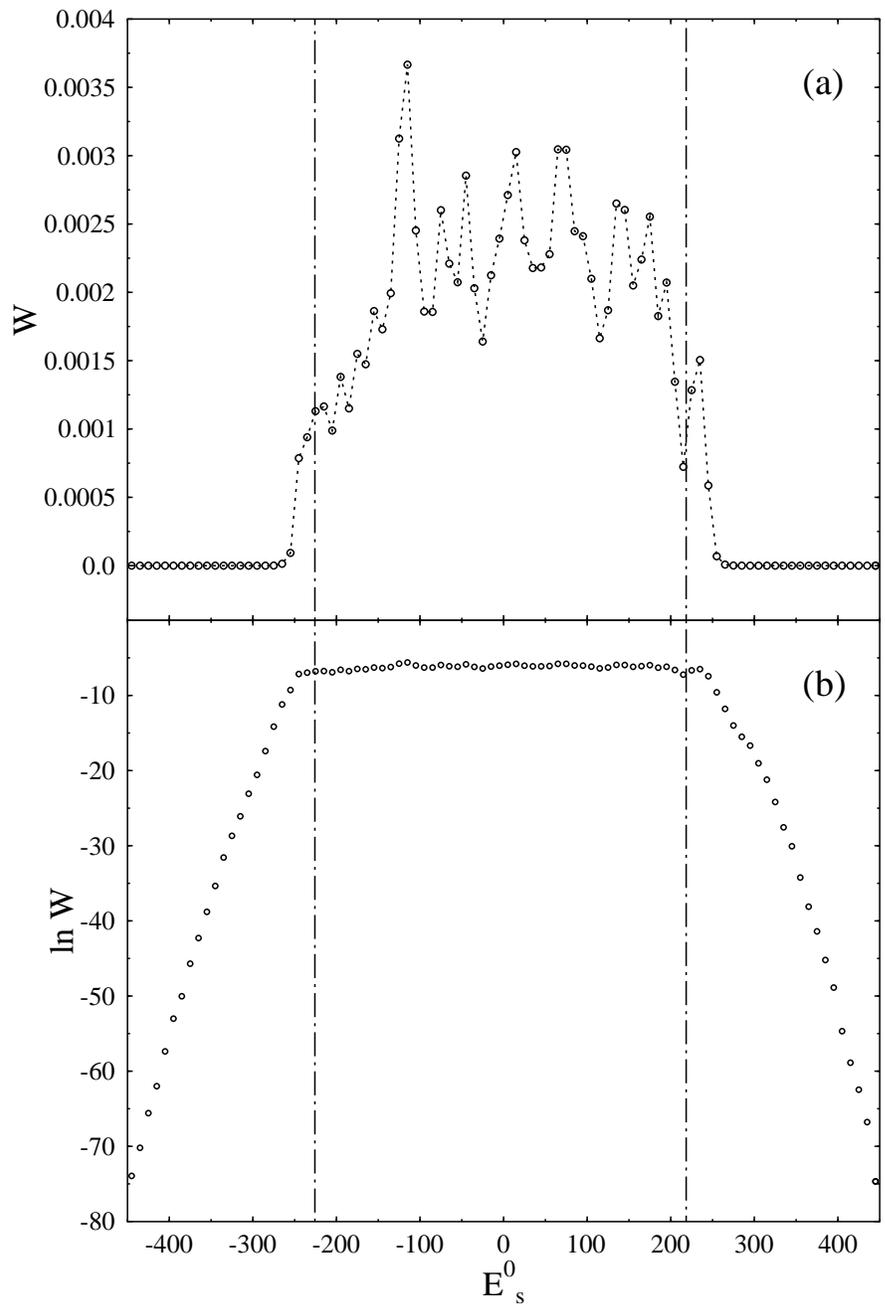
\begin{figure}
\caption{Same as in Fig.~2 for $\lambda =30.0$ and $N=900$. }
\end{figure}

%Fig.10
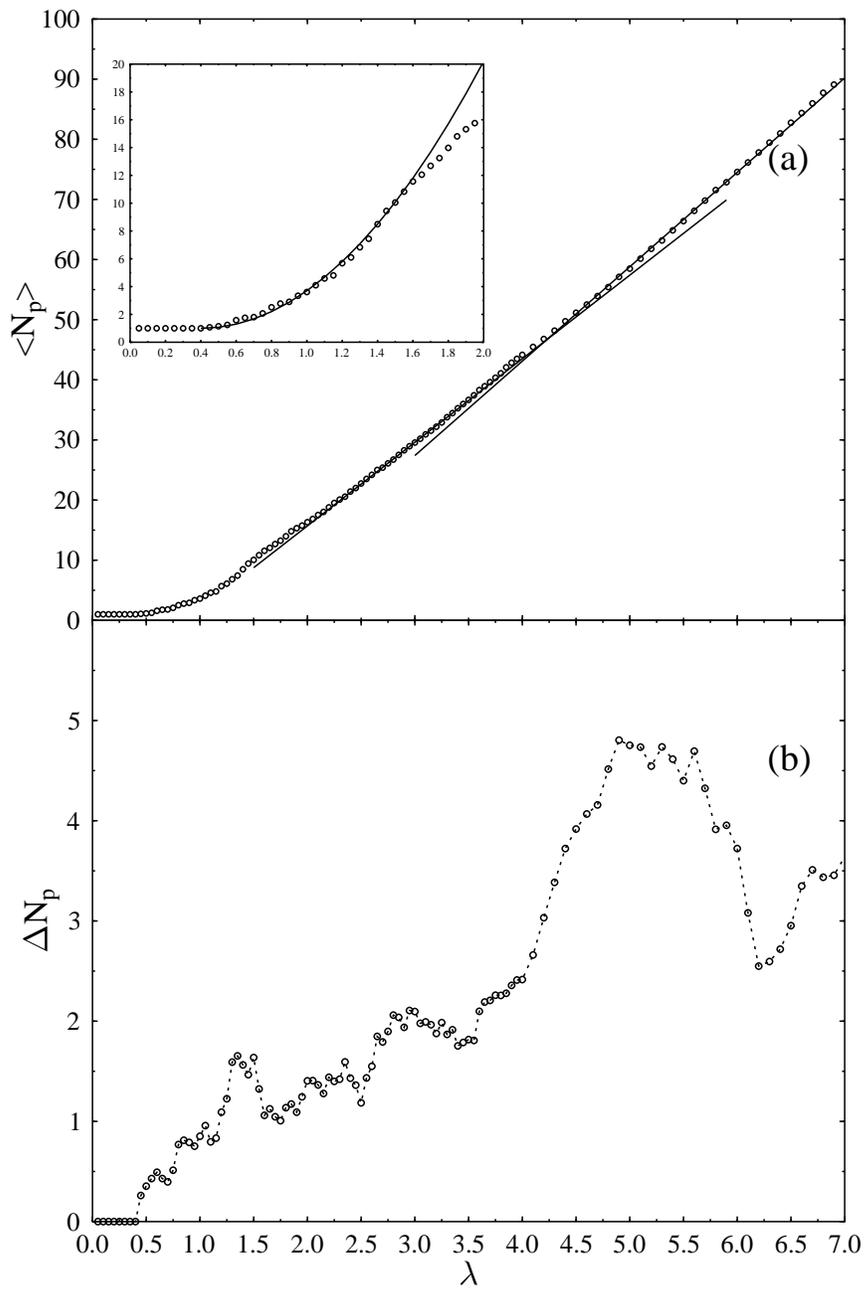
\begin{figure}
\caption{(a) Variation of the average size
$\langle N_p \rangle $ (circles) of  non-perturbative
parts of eigenstates in the middle energy region
with the perturbation strength $\lambda $
for $N=300$. The two solid straight lines are fitting lines.
The
inset shows the fitting curve of the form $7.5(\lambda -0.4)^2+1.0$
for $\lambda $ from 0.4 to 1.5.
(b) Variation of the variance of $N_p$, $\Delta N_p$ (circles),
with $\lambda $. 
}
\end{figure}   

%Fig.11
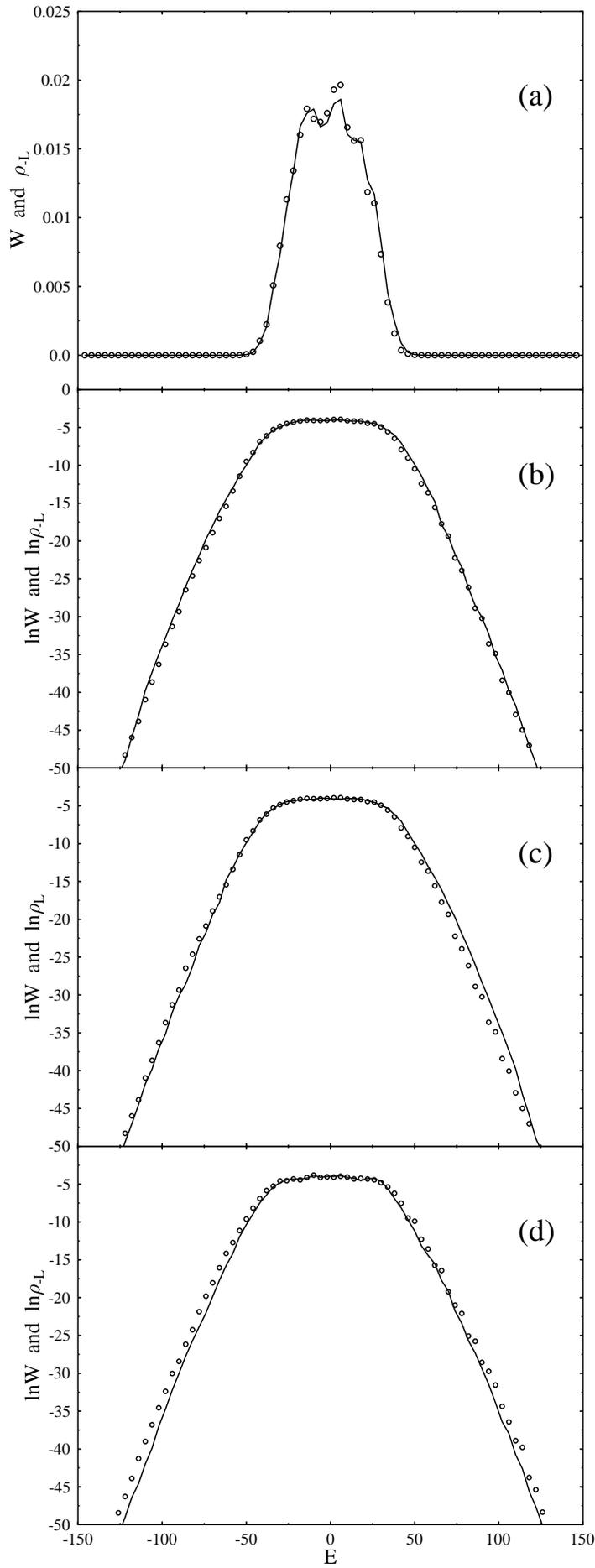
\begin{figure}
\caption{ (a) A comparison between an average shape of eigenfunctions
$W(E)$ (circles) and a related average shape of inverted LDOS
$\rho_{-L}(E)$ (solid curve) for $\lambda =4.0$ and $N=900$. 
(b) Same as in (a) in logarithm scale.
(c) A comparison between the $W(E)$ (circles) in (a) 
and the LDOS $\rho_{L}(E)$ (solid curve) related to the $\rho_{-
L}(E)$ in (a) in logarithm scale.
(d) Same as in (b) for $N=300$. }
\end{figure}   

%Fig.12
\begin{figure}
\caption{ Comparisons between 
the rescaled LDOS $\rho_{L}^{rs}$ (histograms)
obtained when $N=500$ and the prediction of semicircle law
(solid curve) for $\lambda =1$.}
\end{figure}   

%Fig.13
\begin{figure}
\caption{ Same as in Fig.~12 for larger values of $\lambda $. }
\end{figure}

%Fig.14
\begin{figure}
\caption{ The average shape of LDOS $\rho_{L}(E)$ (histograms)
obtained when $N=500$ and the fitting curves (dashed curves)
of the Breit-Wigner form. }
\end{figure}   

%Fig.15
\begin{figure}
\caption{ Same as in Fig.~14 for larger values of $\lambda $. }
\end{figure}

%Fig.16
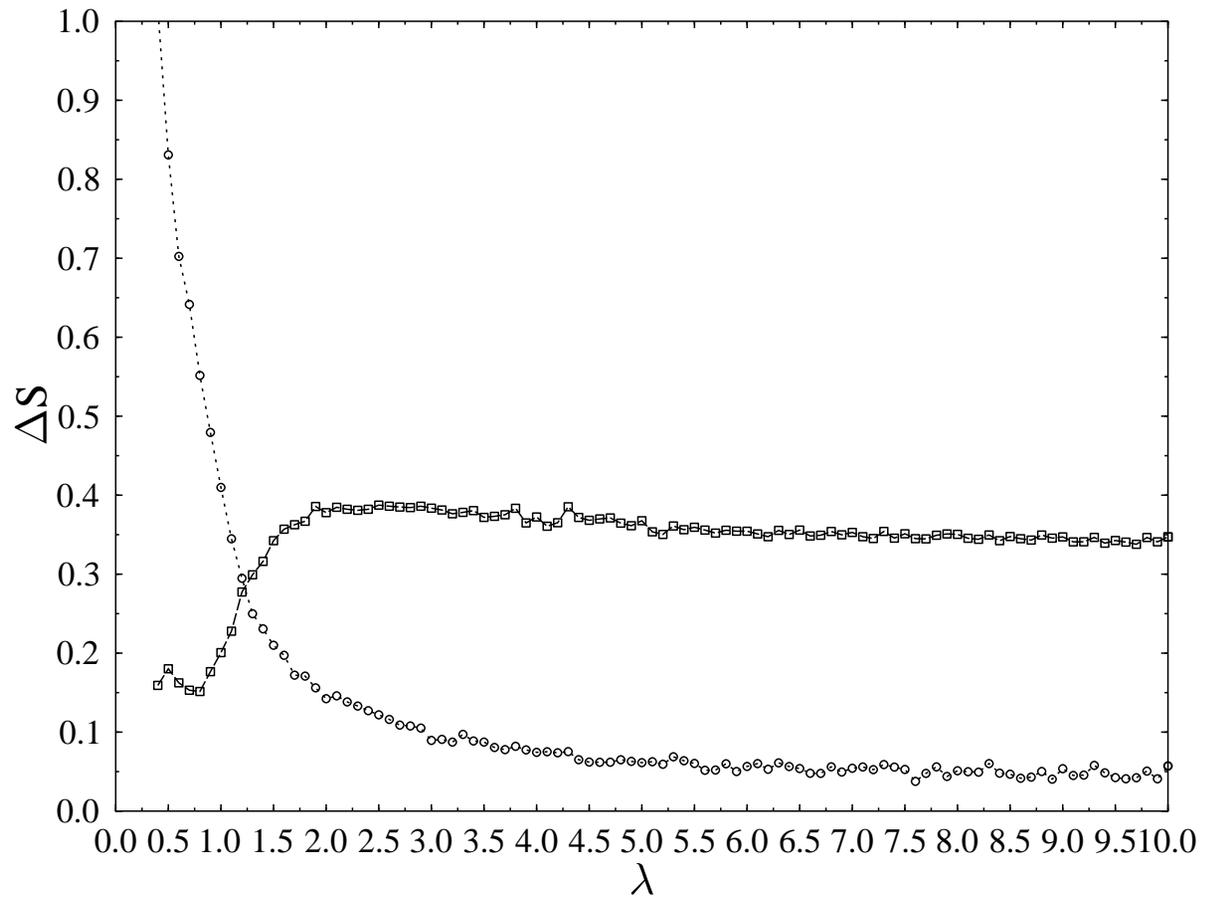
\begin{figure}
\caption{ Variation of the area difference $\Delta S$ (circles)
between the 
rescaled LDOS and the semicircle form for
$\lambda =1$, and variation of the
area difference $\Delta S$ (squares) between the average shape of
LDOS and the fitting Breit-Wigner curves with $\lambda $ 
for $N=500$.}
\end{figure}

%Fig.17
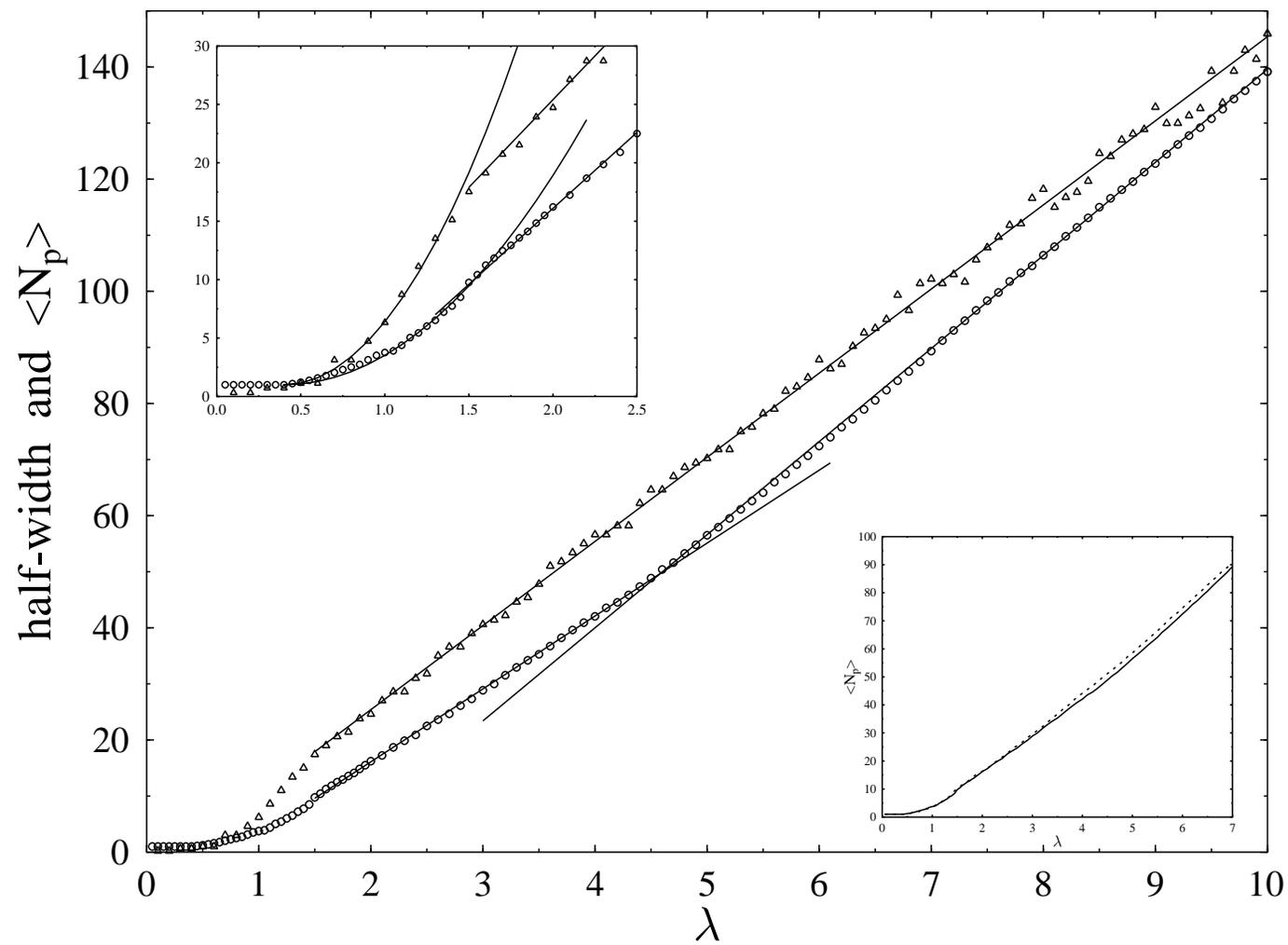
\begin{figure}
\caption{ The triangles show the values of
 the half-width of the average shape of LDOS, and the circles show
 the values of the average size $\langle N_p \rangle $
 of non-perturbative parts of eigenfunctions when $N=500$. The three solid straight
lines are fitting  lines. 
The upper-left inset shows the fitting curve of the form
$(7(\lambda -0.4)^2+1.0)$ for $\langle N_p \rangle $ and the fitting
curve of the form $(15(\lambda -0.4)^2 +1.0)$ for the half-width of the
LDOS for $\lambda $ from 0.4 to 1.5. The lower-right inset gives
a comparison between the $\langle N_p \rangle $ in Fig.~10 obtained when
$N=300$ (dashed curve) and the $\langle N_p \rangle $ here
obtained when $N=500$ (solid curve). }
\end{figure}

%\begin{center} Tables \end{center}

%\begin{table} 
%\begin{tabular}{|c|c|c|c|c|c|c|}
%\begin{tabular}{*{7}{c|}}
%\hline 
%          $\alpha $ & 148 & 149 & 150 & 150 & 151 & 151 \\
%\hline     $p_1$ & 148 & 149 & 149 & 150 & 150 & 151 \\
%\hline     $p_2$ & 148 & 149 & 150 & 151 & 151 & 152 \\
%\hline \end{tabular} 
%\caption{some values of $p_1$ and $p_2$ for the case of %$N=300$, $b=10$ and $\lambda =0.6$. }
%\label{table1} \end{table}

\end{multicols}
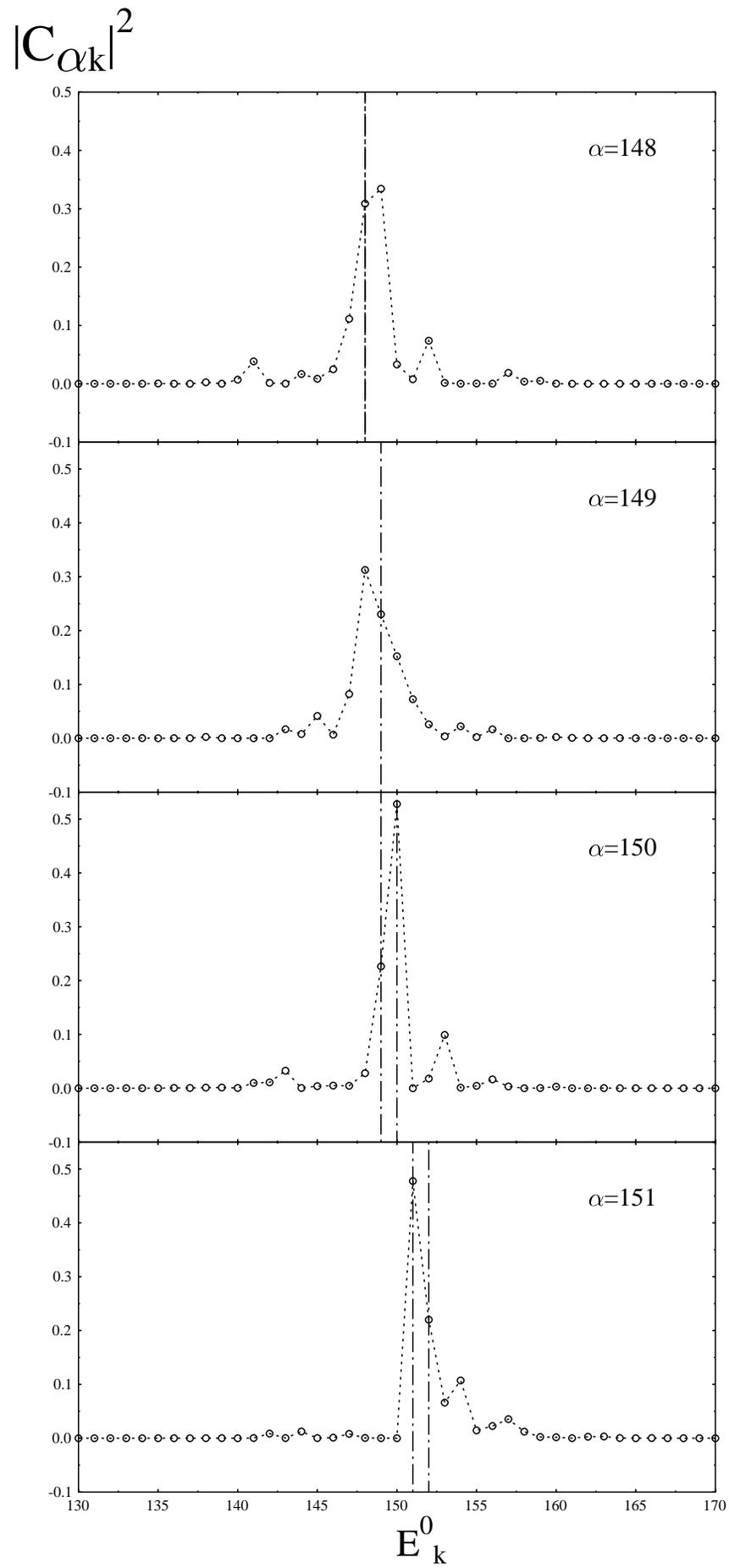
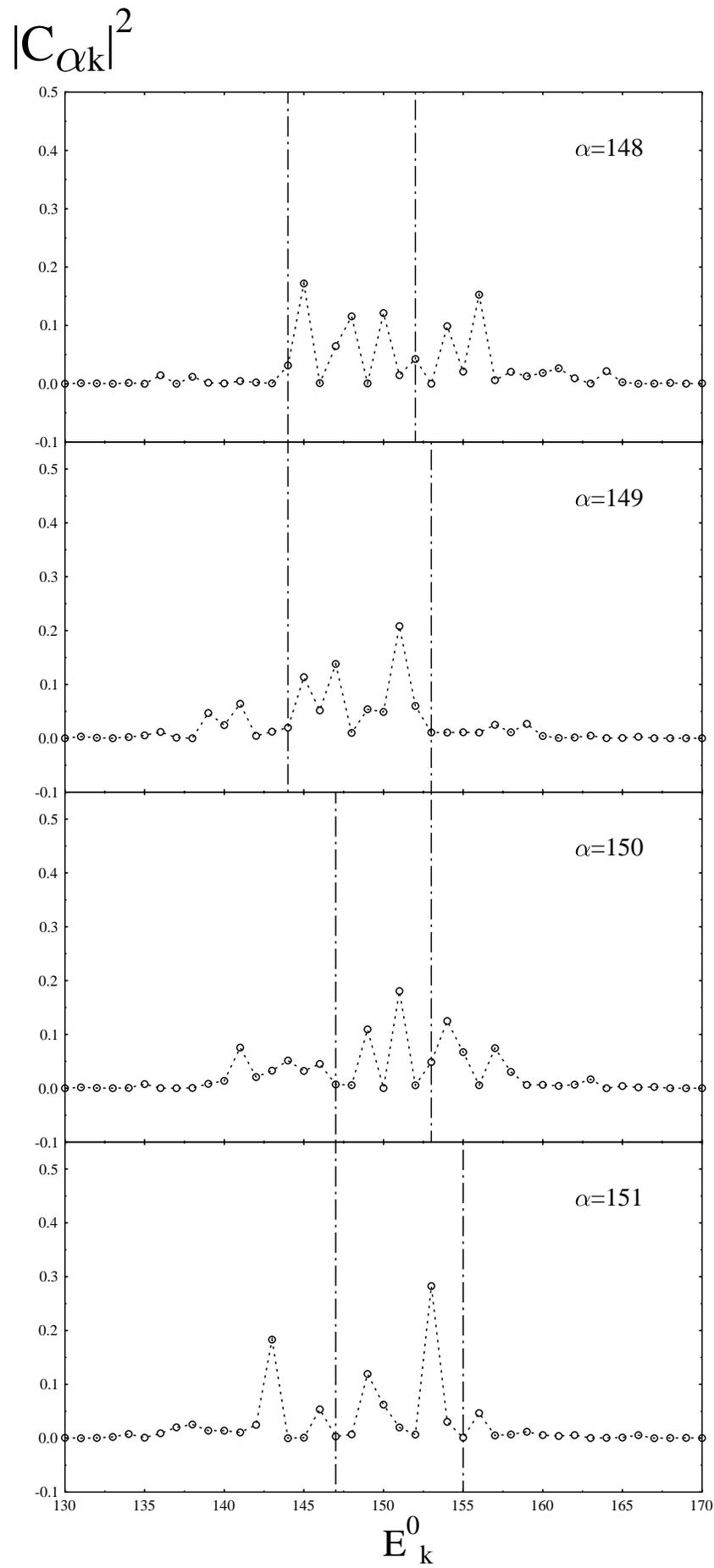
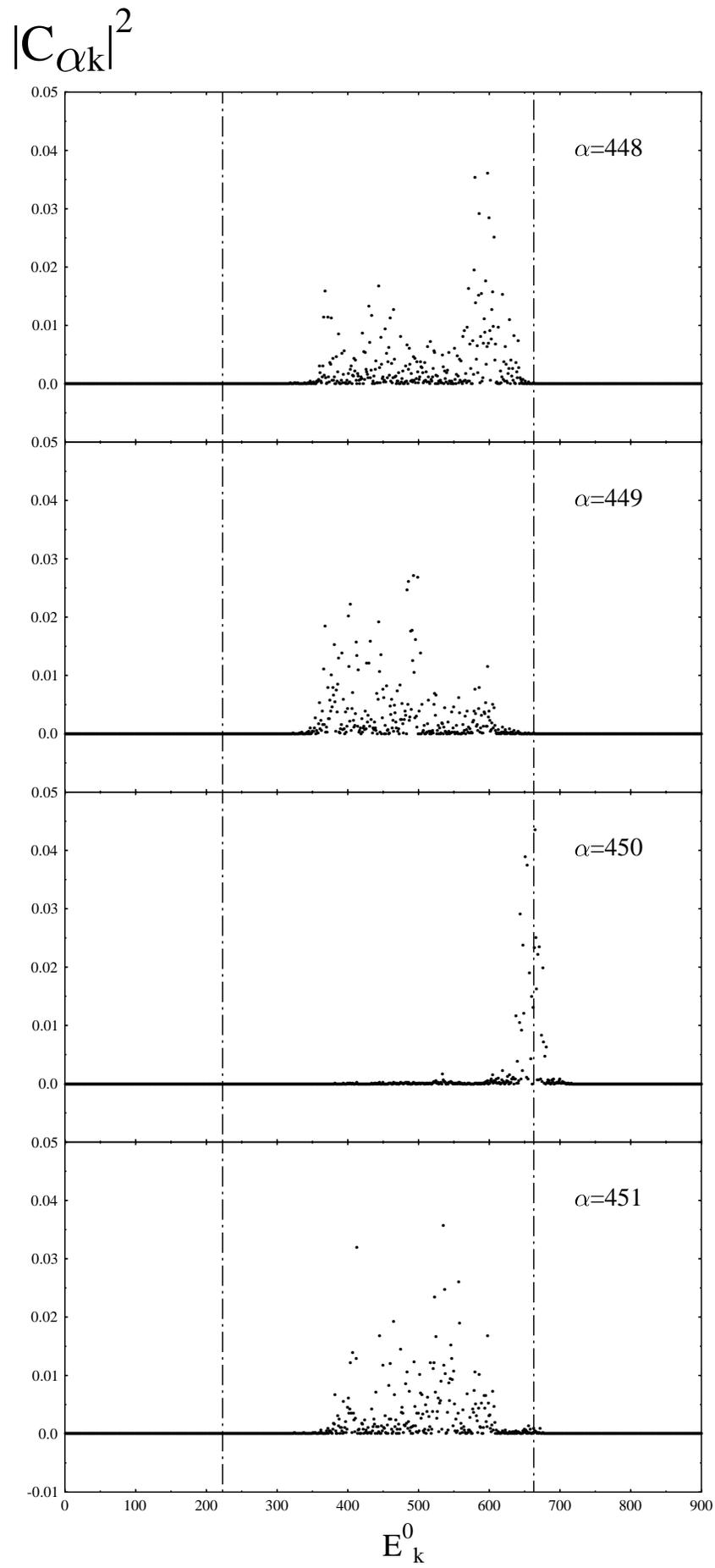
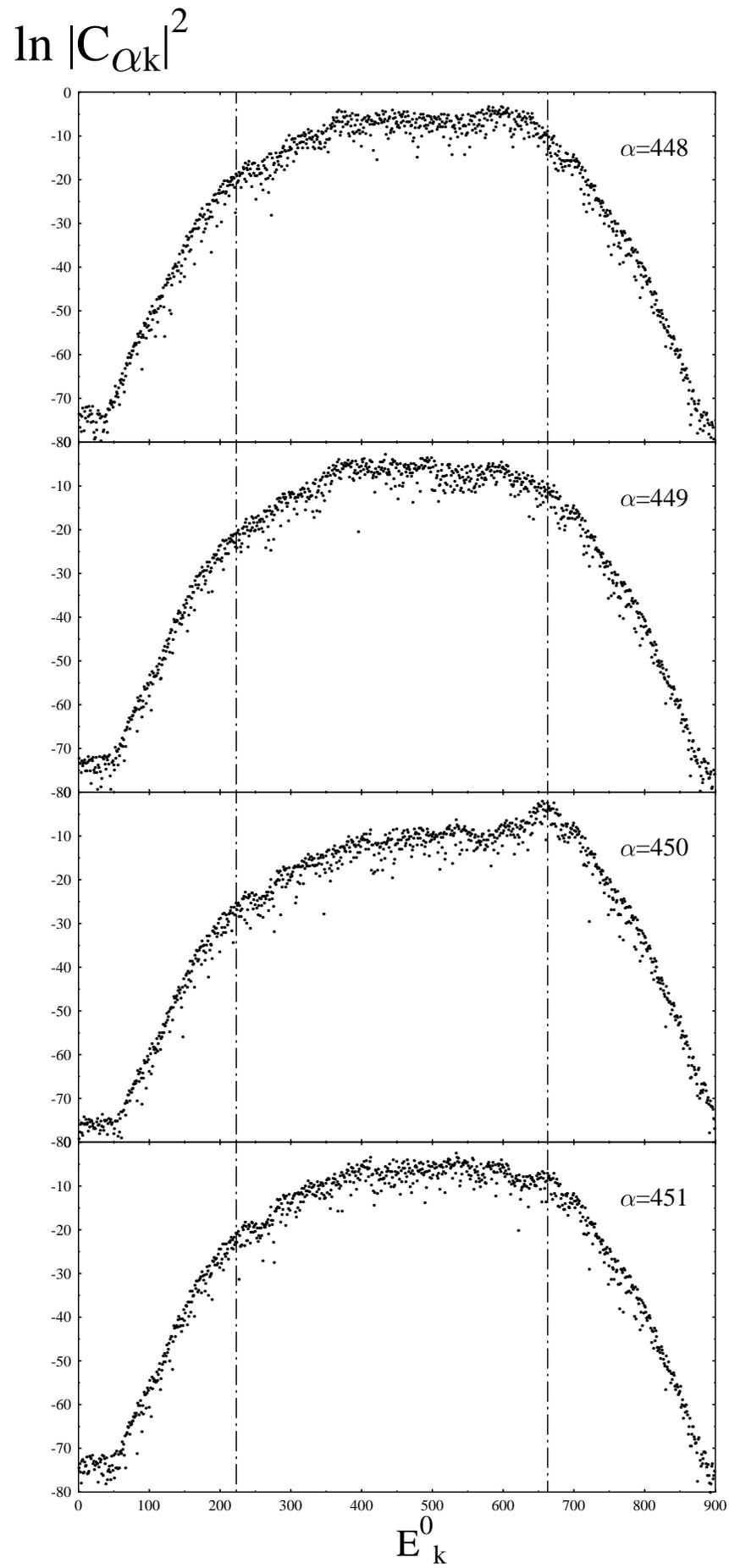
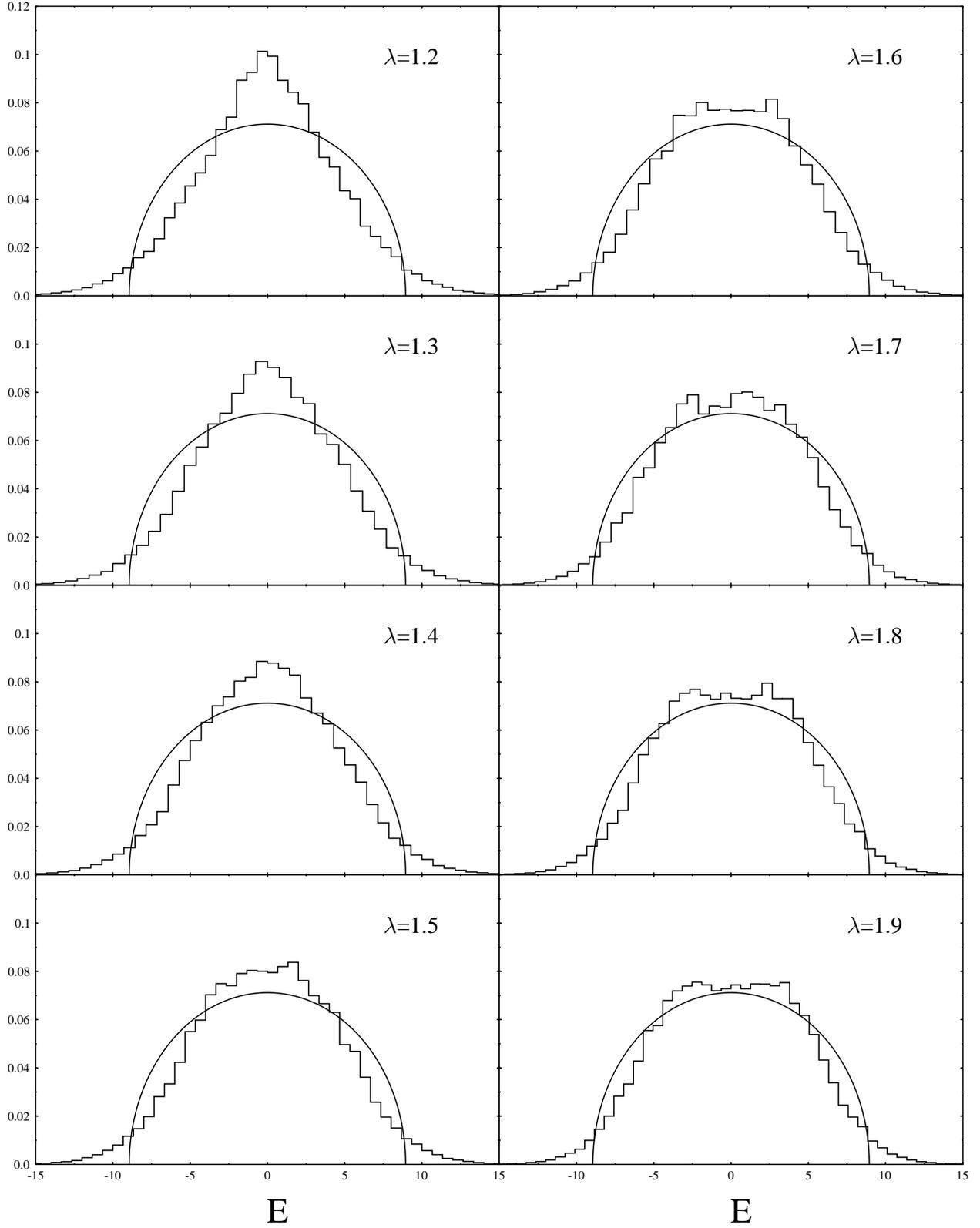
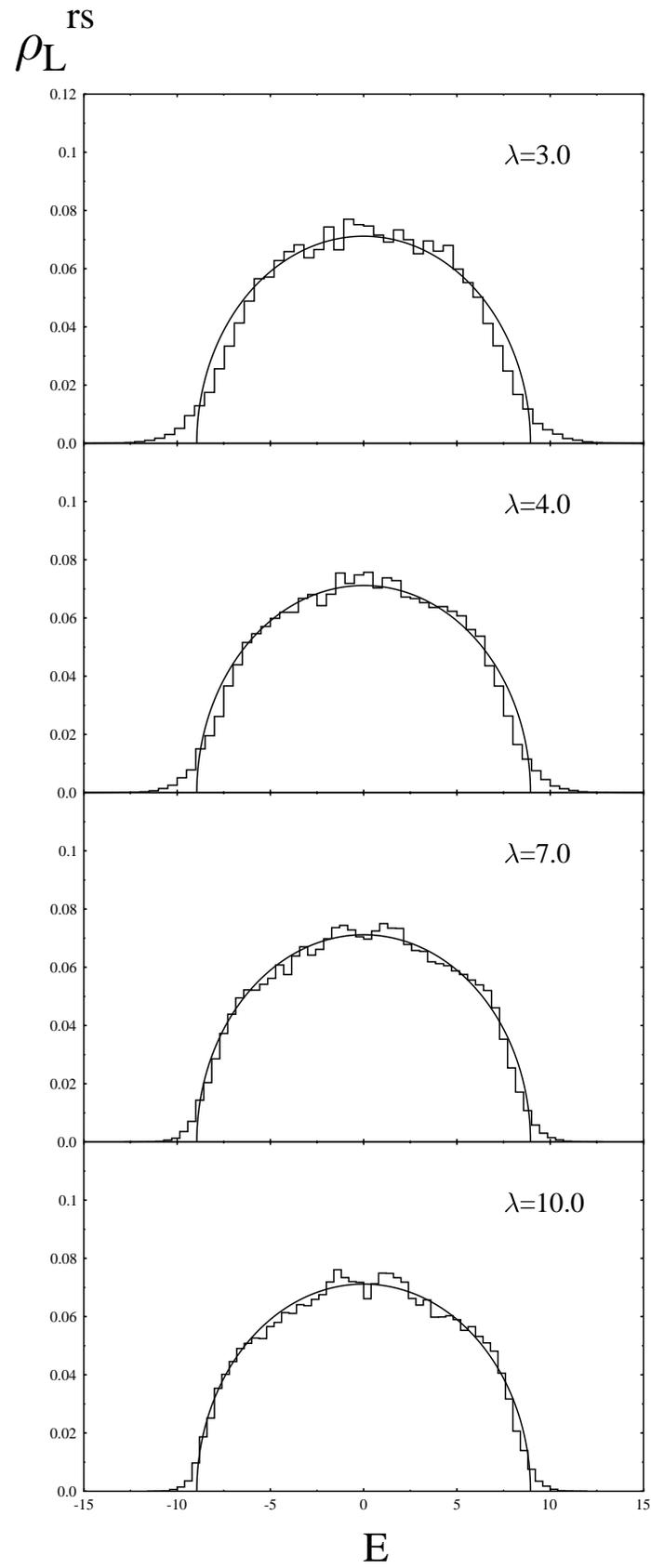
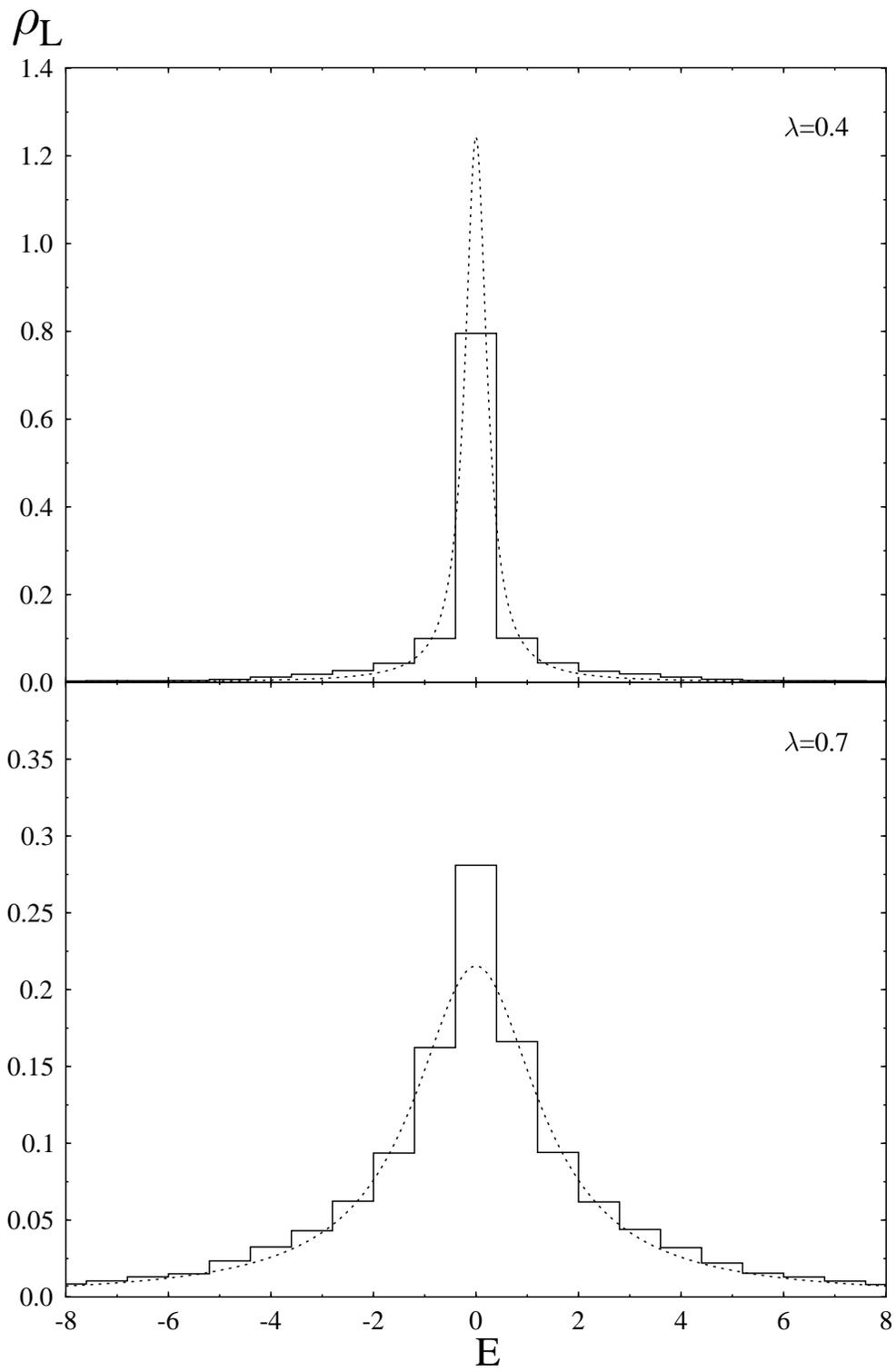
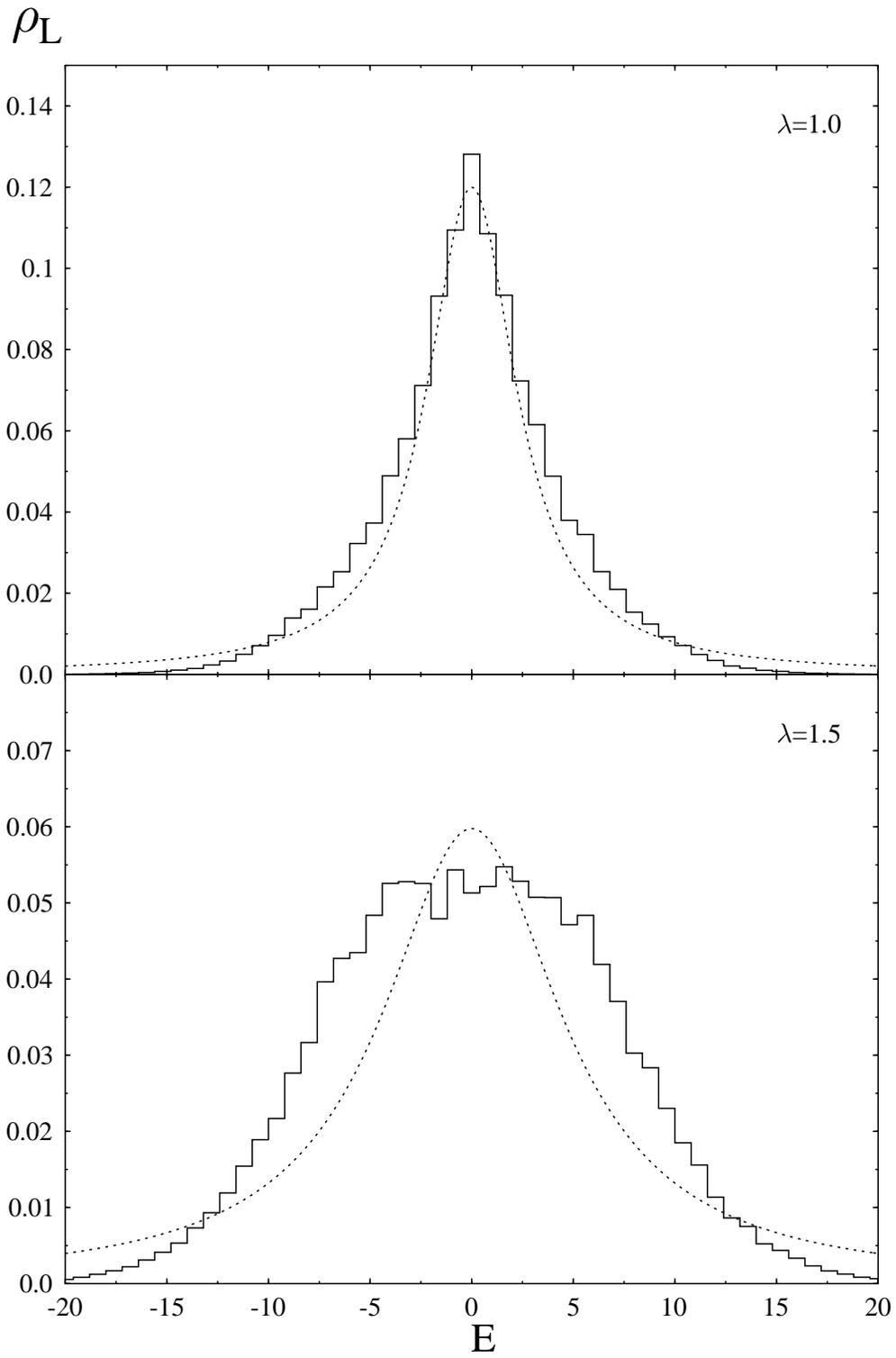
\end{document}